\documentclass{article}

\PassOptionsToPackage{numbers, compress}{natbib}

\usepackage[nonatbib,preprint]{neurips_2023}
\usepackage{authblk}

\usepackage[normalem]{ulem}

\usepackage{setspace}
\usepackage{comment}
\usepackage{amsmath,amsthm,amssymb}
\usepackage{ulem}
\usepackage{makecell}

\newcommand{\bs}{\boldsymbol}
\def\RR{ \mathbb R}
\newcommand{\refeqp}[1]{Equation (\ref{#1})}

\providecommand{\keywords}[1]
{\textbf{\text{Keywords: }} #1}
\newcommand{\ee}{\end{equation}}
\newcommand{\be}{\begin{equation}}
\newcommand{\ec}{\end{center}}
\newcommand{\bc}{\begin{center}}
\newcommand{\eea}{\end{eqnarray}}
\newcommand{\bea}{\begin{eqnarray}}
\newcommand{\bd}{\begin{description}}
\newcommand{\ed}{\end{description}}
\newcommand{\bi}{\begin{itemize}}
\newcommand{\ei}{\end{itemize}}

\newcommand{\bx}{\bs{x}}

\newcommand{\bt}{\bs{\theta}}

\usepackage{commath}
\usepackage{mathtools}
\usepackage{algorithm,algpseudocode}
\usepackage{hyperref}
\usepackage{color}
\usepackage{amsfonts}

\usepackage{diagbox}
\usepackage{bm}
\usepackage{booktabs}       
\usepackage{multirow}
\usepackage{subfigure}
\usepackage[percent]{overpic}
\usepackage{caption}

\newcommand{\ked}{{\kappa_{eff}\in\mathcal{T}_d}}
\newcommand{\bmb}{{\bm{\beta}}}

\newcommand{\bmtheta}{{\bm{\theta}}}

\algnewcommand{\Inputs}[1]{%
  \State \textbf{Inputs:}
  \Statex \hspace*{\algorithmicindent}\parbox[t]{.8\linewidth}{\raggedright #1}
}
\algnewcommand{\Initialize}[1]{%
  \State \textbf{Initialize:}
  \Statex \hspace*{\algorithmicindent}\parbox[t]{.8\linewidth}{\raggedright #1}
}
\algnewcommand{\Outputs}[1]{%
  \State \textbf{Outputs:}
  \Statex \hspace*{\algorithmicindent}\parbox[t]{.8\linewidth}{\raggedright #1}
}

\title{Design-GenNO: A Physics-Informed Generative Model with Neural Operators for Inverse Microstructure Design}

\author[a]{Yaohua Zang}
\author[a,b]{Phaedon-Stelios Koutsourelakis}
\affil[a]{Technical University of Munich, Professorship of Data-driven Materials Modeling, School of Engineering and Design, Boltzmannstr. 15, 85748 Garching, Germany}
\affil[b]{Munich Data Science Institute (MDSI - Core member), Garching, Germany}
\affil[ ]{\text{\{yaohua.zang, p.s.koutsourelakis\}@tum.de}}
\begin{document}
\maketitle
\begin{abstract}
Inverse microstructure design plays a central role in materials discovery, yet remains challenging due to the complexity of structure–property linkages and the scarcity of labeled training data. We propose Design-GenNO, a physics-informed generative neural operator framework that unifies generative modeling with operator learning to address these challenges. In Design-GenNO, microstructures are encoded into a low-dimensional, well-structured latent space, which serves as the generator for both reconstructing microstructures and predicting solution fields of governing PDEs. MultiONet-based decoders enable functional mappings from latent variables to both microstructures and full PDE solution fields, allowing a multitude of design objectives to be addressed without retraining. A normalizing flow prior regularizes the latent space, facilitating efficient sampling and robust gradient-based optimization. A distinctive feature of the framework is its physics-informed training strategy: by embedding PDE residuals directly into the learning objective, Design-GenNO significantly reduces reliance on labeled datasets and can even operate in a self-supervised setting.
We validate the method on a suite of inverse design tasks in two-phase materials, including effective property matching, recovery of microstructures from sparse field measurements, and maximization of conductivity ratios. Across all tasks, Design-GenNO achieves high accuracy, generates diverse and physically meaningful designs, and consistently outperforms the state-of-the-art method. Moreover, it demonstrates strong extrapolation by producing microstructures with effective properties beyond the training distribution. These results establish Design-GenNO as a robust and general framework for physics-informed inverse design, offering a promising pathway toward accelerated materials discovery.
\end{abstract}
\section{Introduction}
\label{sec:introduction}
Recent advances in manufacturing technologies—such as additive manufacturing, 3D printing, and real-time process control—have made it possible to precisely engineer the spatial distribution of material constituents, i.e., the microstructure. This capability enables microstructure-centered design, where optimal configurations are sought to achieve target macroscopic properties, with applications ranging from structural alloys \cite{nie1998microstructural}, aerospace composites \cite{kassapoglou2010design}, and biomedical implants \cite{alabort2022alloys} to energy systems \cite{feng2022machine} and materials with tailored porosity or thermal behavior \cite{zang2025psp,zhou2008computational}.
In contrast to forward modeling, which predicts effective properties from known structures, inverse design addresses the far more challenging problem of determining microstructures that meet prescribed property requirements. This often involves navigating high-dimensional, nonlinear, and multiscale design spaces. 
Traditionally, these problems have been approached through experimental trial-and-error or brute-force computational searches—methods that are expensive, time-consuming, and often impractical for complex systems. These limitations highlight the critical need for more efficient, scalable inverse design frameworks that integrate physics-based insights with advanced computational tools to accelerate material discovery and innovation.

Over the past decade, significant efforts have been devoted to tackling the challenges of inverse microstructure design \cite{fullwood2010microstructure,zunger2018inverse}. Among the most widely adopted strategies are iterative optimization approaches, such as gradient-based topology optimization \cite{christiansen2021inverse,feng2024topology}, genetic algorithms \cite{kulkarni2004microstructural,athinarayanarao2023computational}, and simulated annealing \cite{oliveri2020inverse}. These approaches typically operate by repeatedly evaluating the forward structure–property (SP) map and updating candidate microstructures to minimize the discrepancy between predicted and target properties. While effective in certain settings, these methods are often time-consuming and computationally intensive due to the large number of forward simulations required. Moreover, they are prone to convergence issues such as getting trapped in local minima—particularly in high-dimensional, irregular design spaces. 
To mitigate these issues, more recent optimization approaches have introduced surrogate-assisted and reduced-order modeling approaches that accelerate the design process. These include models based on Bayesian optimization \cite{honarmandi2022accelerated,generale2023bayesian}, low-dimensional latent variable representations \cite{jung2020microstructure,wang2021data}, and microstructure descriptors \cite{kalidindi2011microstructure,xu2017novel} that simplify the optimization landscape. While such methods can substantially reduce computational cost and improve convergence, they frequently rely on handcrafted features, limited or task-specific training data, and approximations that may not fully capture the complex physics governing SP relationships. Consequently, their generalizability to new material systems or design requirements remains limited.

More recently, deep learning has shown strong potential for addressing challenges in inverse microstructure design \cite{wang2022inverse,zheng2023deep}. In particular, deep generative models (DGMs) such as VAEs \cite{zang2025psp,kim2021exploration,attari2023towards}, GANs \cite{yang2018microstructural,lee2021fast}, flow-based models \cite{yang2023normalizing,mirzaee2025inverse}, and diffusion models \cite{vlassis2023denoising,park2024inverse} have excelled at learning microstructure distributions and generating realistic designs. 
By encoding high-dimensional structures into low-dimensional latent spaces with powerful network architectures like convolutional neural networks (CNNs), these models enable efficient exploration of the design space via search or optimization in the latent space. Therefore, they substantially accelerate the discovery of microstructures with target properties. However, unconditional generation followed by post hoc property filtering \cite{yang2018microstructural,cang2018improving,attari2023towards} is inefficient and unreliable, as it lacks direct control over the generated properties and may result in a large proportion of non-optimal or invalid microstructures. More advanced frameworks have coupled DGMs with property regressors \cite{kim2021exploration,lee2021fast,wang2020deep}, improving controllability and design accuracy but suffering from error accumulation, limited generalization, and reliance on expert knowledge to guide optimization. 
To further enhance control and design fidelity, conditional generative models \cite{yang2023normalizing,mirzaee2025inverse,vlassis2023denoising,park2024inverse} incorporate desired properties directly, providing better controllability and efficiency. Yet, all current DGM-based methods require large labeled datasets comprising paired microstructures and corresponding properties. In practice, such labeled datasets are often scarce or prohibitively expensive to obtain due to the high cost of experiments or high-fidelity numerical simulations. Consequently, there is a pressing need to develop data-efficient, physics-informed DGM approaches.

In parallel, physics-informed machine learning (PIML) has rapidly emerged as a powerful paradigm for scientific computing, with successful applications across a range of disciplines, including numerical PDEs \cite{yu2018deep,sirignano2018dgm,raissi2019physics,grigo_physics-aware_2019,zang2020weak,zang2023particlewnn}, medical imaging \cite{bao2020numerical,scholz2025weak}, fluid dynamics \cite{cai2021physics}, solid mechanics \cite{haghighat2021physics,jin2023recent}, and materials science \cite{kats2022physics,farrag2025physics}. 
By embedding governing physics laws into neural network loss functions, PIML enables physically consistent learning with limited labeled data. 
In the context of inverse design, PIML has been used to construct surrogate models that capture the SP linkages governed by underlying physical laws. For example, a physics-informed neural network (PINN) with hard constraints is developed in \cite{lu2021physics} to solve topology optimization problems, designing geometries that yield a targeted transmitted-wave pattern. In \cite{kumar2022physics}, PINN has also been used to recover optimal diffusion coefficients in NiCoFeCr multi-principal element alloys. In another example, \cite{hasan2024microstructure} proposes a physics-informed LSTM network as a surrogate for predicting microstructural texture evolution across time steps during deformation, demonstrating improved efficiency over traditional CNN-based surrogates in microstructure-sensitive design.
Despite their promise, most PIML methods rely on simplified parametric representations of microstructures, assume full-field targets, or require costly PDE-constrained optimization, making them less suitable for high-dimensional, nonparametric microstructure design problems.

Unlike PIML approaches, which parameterize solution fields directly via neural networks, deep neural operators (DNOs) aim to learn mappings between infinite-dimensional function spaces, i.e., from input functions (e.g., material geometries, loading conditions) to output functions (e.g., fields of interest such as stress, strain, or temperature). This formulation enables DNOs to approximate entire solution operators rather than specific solution instances, offering a promising alternative for inverse microstructure design.
Recent studies have demonstrated the promise of DNOs in inverse design. For example, a multifidelity DeepONet model was introduced in \cite{lu2022multifidelity} to solve inverse design problems in nanoscale heat transport. By integrating with genetic algorithms or topology optimization, this model enables rapid solution of the phonon Boltzmann transport equation (BTE) during design. 
Similarly, the DeepONet architecture has also been employed to capture the relationship between microstructures and their mechanical responses, thereby supporting the inverse design of micro-architected materials with desired properties \cite{jin2025characterization}.
While these models offer benefits such as strong generalization ability, nonlinear operator approximation, and compatibility with multifidelity data, they remain fundamentally data-driven, typically lacking explicit physics integration and latent generative modeling, which hampers efficient exploration of high-dimensional and complex microstructural spaces.
To overcome these limitations, the recently proposed Deep Generative Neural Operators (DGenNO) \cite{zang2025dgenno} framework unifies operator learning with DGMs to solve parametric PDEs and corresponding inverse problems in a physics-informed manner. It integrates a generative model with a neural operator backbone to jointly learn a latent representation of the input functions and the associated solution operator. This fusion enables the framework to model complex input distributions, enforce property constraints through generative sampling, and generalize across varying input conditions. The resulting model not only benefits from the generalization power of operator learning but also inherits the expressiveness and design flexibility of DGMs, making it a promising approach for high-dimensional, microstructure-centered inverse design.

In this work, we propose a novel physics-informed generative neural operator framework, called Design-GenNO, for solving the inverse microstructure design problem, built upon the DGenNO model. Specifically, microstructures are first encoded into a low-dimensional, structured latent space via an encoder network. This latent representation serves as a generator for both reconstructing microstructures and predicting the solution fields of the underlying PDEs that govern the SP linkage. 
To accomplish this, we employ the MultiONet architecture \cite{zang2025dgenno}, which has demonstrated superior approximation capabilities compared to DeepONet. MultiONet is used to construct two decoders: one for reconstructing the microstructure from the latent code, and another for learning the nonlinear operator that maps latent representations to solution fields.
Since the MultiONet operator is fully differentiable, we can efficiently solve the inverse design problem using gradient-based optimization in the latent space. Once the optimal latent vector is found for a given target, the corresponding microstructure can be directly generated using the trained decoder.
A key feature of our framework is its physics-informed training strategy, which incorporates the governing PDEs directly into the loss function through residual minimization (either in strong or weak form). This allows the model to be trained without requiring extensive labeled input–output pairs, enabling data-efficient or even self-supervised learning.
To further enhance generative performance and improve inverse design efficiency, we incorporate a normalizing flow model that provides a structured, non-Gaussian, and potentially multimodal, learnable prior in the latent space, which increases the expressivity and robustness of the proposed framework \cite{xu_necessity_2019}.  
In summary, the main contributions of this work are:
\begin{itemize}
    \item A unified generative neural operator framework, named Design-GenNO, is developed for inverse microstructure design, combining DGMs with the function-to-function mapping power of DNOs. This integration enables efficient inverse design over the latent space with gradient-based optimization. Moreover, a normalizing flow model is introduced to provide a good prior for the optimization. 
    \item A novel physics-informed training paradigm that embeds governing PDEs into the learning process via residual minimization. This strategy significantly reduces the reliance on labeled data and allows for effective training with partially or fully unlabeled datasets while ensuring physical consistency.
    \item Comprehensive validation on inverse design tasks involving two-phase materials with both effective property targets and spatially distributed response field targets. Numerical results show strong interpolation ability of the proposed framework, which successfully generates microstructures with properties not present in the training data.
\end{itemize}

The remainder of this paper is structured as follows. Section \ref{sec:problem} formulates the microstructure-centered inverse design problems, outlining different design objectives and the governing SP linkage. Section \ref{sec:method} introduces our proposed physics-informed deep generative neural operator framework, including the model details, physics-informed training strategy, and its application to inverse design problems with different targets. Section \ref{sec:experiments} demonstrates the effectiveness of the approach across multiple design tasks and benchmarks it against state-of-the-art methods. Finally, Section \ref{sec:conclusion} summarizes the main findings and discusses potential directions for future work.

\section{Problem Statement}
\label{sec:problem}
\subsection{Microstructure-centered inverse design problems}
\label{sec:defp13}
Microstructure-centered inverse design is a fundamental challenge in computational materials science. It aims to discover spatial arrangements of constituent phases, i.e., microstructures, that yield desired properties, such as effective mechanical properties or spatially distributed response fields. 
In this work, we focus on two-phase composite materials defined on a bounded spatial domain $\Omega \subset \mathbb{R}^2$, discretized into a $K \times K$ uniform grid. Each pixel or spatial location $\bm{x} \in \Omega$ is assigned a local property value via a coefficient function $\mu(\bm{x}): \Omega \rightarrow {\kappa_1, \kappa_2}$, where $\kappa_1$ and $\kappa_2$ represent the property corresponding to phase 1 and phase 2, respectively. The microstructure $\mu$ is therefore modeled as a piecewise-constant field that encodes the material heterogeneity.

We consider two categories of properties that serve as inverse design targets: (1) \textbf{macroscopic effective properties}, and (2) \textbf{microscopic field responses}. Specifically, the effective property is denoted by $\kappa_{\text{eff}}$. The field response (e.g., displacement field or temperature field) is characterized by the spatial distribution $u(\bm{x})$.
To demonstrate the effectiveness of the proposed method, we focus on the following three representative inverse design problems:
\begin{itemize}
    \item \textbf{P1}: Given a target region $\mathcal{T}_d$ in the effective property space, design microstructures $\mu$ whose effective property $\kappa_{\text{eff}}$ lies within $ \mathcal{T}_d$.
    \item \textbf{P2}: Given a target field $u_d$ on $\Omega$, identify  microstructure(s) $\mu$ such that the corresponding field prediction satisfies $u \approx u_d$.
    \item \textbf{P3}: Given a utility function $F$, design microstructures $\mu$ that maximize the objective $F(\kappa_{eff})$.
\end{itemize}

\subsection{The structure–property linkage}
\label{sec:SP-link}
The forward problem, also referred to as the Structure-Property (SP) linkage, maps microstructures to physical responses via the solution of a PDE system. Without loss of generality, a general PDE can be written as:
\begin{equation}\label{eq:pde_general}
\begin{array}{ll}
\mathcal{N}(u, \mu) = s, \quad \bm{x}\in\Omega \subset \RR^d,\\
\mathcal{B}(u) = g, \quad \bm{x}\in\partial\Omega,
\end{array}
\end{equation}
where $\mu$ denotes the coefficient function characterizing the microstructure, $u$ is the PDE solution (field response), $s$ is the source term, and $g$ is the boundary condition. Here, $\mathcal{N}$ and $\mathcal{B}$ represent the differential and boundary operators, respectively.
In problem \textbf{(P2)}, the forward task corresponds to solving the PDE \eqref{eq:pde_general} directly, as the design target is the PDE solution $u$. In problems \textbf{(P1)} and \textbf{(P3)}, however, the design target is the effective property $\kappa_{eff}$ or its function $F(\kappa_{eff})$, both of which depend on the PDE solution $u$ and, potentially, the microstructure $\mu$. Thus, in these cases, the SP linkage involves not only solving the PDE but also applying a compression map to extract the effective property $\kappa_{eff}$ from $u$ (and possibly $\mu$).

\subsection{Challenges in inverse microstructure design}
Solving the inverse microstructure design problem is inherently challenging due to a combination of factors. First, the design space is high-dimensional and discrete: for a 2D domain, each microstructure $\mu$ is represented as a $K \times K$ binary image, resulting in $2^{K\times K}$ possible configurations. Small changes to the microstructure, such as flipping a single pixel, can cause large changes in the effective properties, making the design space combinatorially complex and highly non-linear. These changes in properties are amplified as the contrast between the properties of the constitutive phases becomes larger \cite{torquato_random_2002}.
Second, the piecewise-constant nature of microstructures makes gradient-based optimization particularly difficult. Since $\mu$ is discontinuous, computing gradients with respect to the design variables is not directly feasible. A common solution used in topology optimization is to relax the design representation by approximating microstructures with smooth, continuous functions. While this enables gradient computation, it fundamentally alters the design space that may not correspond to physically realizable microstructures. Moreover, relaxation methods suffer from difficulties in scaling to high-dimensional design domains, often requiring additional post-processing (e.g., thresholding) that can distort the optimized design, introduce artificial artifacts, or push the solution away from the true optima.
Third, the problem is often ill-posed, particularly in effective property matching \textbf{(P1)} and utility function maximization \textbf{(P3)}, as many distinct microstructures can yield the same macroscopic behavior, making the inverse mapping one-to-many and necessitating probabilistic or generative frameworks. 
Moreover, data scarcity poses a practical limitation: obtaining labeled microstructure–property pairs requires repeated numerical PDE solves, which are computationally intensive and costly, especially when high-resolution simulations are needed. 
Finally, the dependence on forward PDE evaluations during optimization introduces significant computational overhead, which becomes a bottleneck for large-scale or real-time inverse design unless efficient surrogate models are available. These challenges collectively motivate the development of a data-efficient, physics-informed generative approach for inverse microstructure design.

\section{Methodology}
\label{sec:method}
In this section, we present a physics-informed framework for inverse microstructure design, called Design-GenNO, which is centered around the DGenNO formulation \cite{zang2025dgenno} — a model that combines deep generative modeling with neural operator learning to capture the complex SP relationship. We begin by describing the different modalities in the training datasets, which can include unlabeled, labeled, and {\em virtually} labeled data, as we explain in the sequel. We then detail the DGenNO architecture and its physics-informed training procedure. Next, we introduce a novel extension involving a normalizing flow model that learns a flexible prior over the latent space. Finally, we summarize the complete Design-GenNO framework and describe how it enables efficient and physically consistent microstructure design by optimizing property-related targets.

\subsection{Training Data Modalities}
\label{sec:training_data}
Training data-driven models has traditionally relied on simulation data (i.e., pairs of PDE inputs and outputs) and, more recently, on physics-informed signals. Our framework can additionally leverage unlabeled data (i.e., PDE inputs without corresponding outputs), which have received considerably less attention in the literature~\cite{rixner_probabilistic_2021} despite being significantly cheaper to obtain than labeled data, both experimentally and computationally.
In the following, we denote data or observables with a hat $\hat{}$ and explain their meaning in the sequel.

The three training modalities considered are:
\bi
\item \textbf{Unlabeled data} $\mathcal{D}_u=\{ \hat{\mu}^{(n_u)} \}_{n_u=1}^{N_u}$.  
    In practice, each $\hat{\mu}^{(n_u)}$ is represented as a $K^d$ ($K \gg 1$) binary or grayscale image, placing the design in a high-dimensional and discrete space that encodes the corresponding microstructure over the problem domain $\Omega \subset \mathbb{R}^d$. 
    \item \textbf{Labeled data pairs} $\mathcal{D}_l=\{ \hat{\mu}^{(n_l)}, \hat{\bm{u}}^{(n_l)} \}_{n_l=1}^{N_l}$.  
    These are typically obtained either experimentally or numerically by solving the governing PDE  in \refeqp{eq:pde_general}. While such data pairs are highly informative, they are also the most computationally expensive to obtain. In practice, $N_l \ll N_u$, and the smaller $N_l$ is (ideally $N_l = 0$), the lower the training cost becomes. In general, they pertain to noisy values of the PDE-output (e.g., at some grid points $\Xi_{u}\subset\Omega$) which can be  integrated with a likelihood of the form:
    \be
    p(\hat{\bm{u}}^{(n_l)} |u^{(n_l)}) = \mathcal{N}(\hat{\bm{u}}^{(n_l)} | o(u^{(n_l)}), \lambda_{data}^{-1} \bs{I})
    \label{eq:ulike}
    \ee
    where $o(.)$ denotes the observation filter (e.g., $o(u^{(n_l)})=u^{(n_l)}(\Xi_u)$), $u^{(n_l)}$ the true but unknown PDE-solution field and $\lambda_{data}^{-1}$ the variance of the observation noise.
   \item \textbf{Virtual data pairs} $\mathcal{D}_v=\{ \hat{\mu}^{(n_v)}, \hat{\bs{R}}^{(n_v)} \}_{n_v=1}^{N_v}$. These are based on the concept of {\em virtual observables} \cite{kaltenbach_incorporating_2020}, which has been successfully used in order to incorporate physical information into probabilistic models \cite{chatzopoulos2024physics,scholz2025weak}. For the sake of illustration in place of \refeqp{eq:pde_general}, we consider 
   the elliptic operator $\mathcal{N}(u,\mu) = -\nabla \cdot (\mu \nabla u)$ and $M$ weighted residuals associated with a test function $w_m \in H_0^1(\Omega), m=1,\ldots M$ \footnote{To maintain flexibility for complex geometries and avoid dense quadrature requirements, we use compactly supported radial basis functions (CSRBFs)~\cite{zang2023particlewnn} as test functions.}:
    \be\label{eq:weak_residual}
    r_m(\mu,u) := \int_\Omega \Big(\mu(\bm{x}) \nabla u(\bm{x}) \cdot \nabla w_m(\bm{x}) - s(\bm{x}) w_m(\bm{x})\Big)\, d\bm{x}, \qquad \forall m
    \ee
    For a solution pair $(\mu,u)$ of the PDE, these residuals would be equal to 0. For a given PDE-input $\hat{\mu}^{(n_v)}$, we assume  that we have {\em virtually} observed that their   values are equal to $0$ and denote the corresponding vector with $\hat{\bs{R}}^{(n_v)}=\bs{0}= \{\hat{r}^{(n_v)}_1,\dots,\hat{r}^{(n_v)}_M\}$. These can give rise to a {\em virtual} likelihood of the form \cite{zang2025dgenno}:
    \be
    p(\hat{\bs{R}}^{(n_v)} | u^{(n_v)}, \hat{\mu}^{(n_v)})\propto \prod_{m=1}^{M} \exp\{-\frac{\lambda_{\mathrm{pde}}}{2} r_m^2(\hat{\mu}^{(n_v)},u^{(n_v)}) \}
    \label{eq:vlike}
    \ee
    where $\lambda_{\mathrm{pde}}$ controls the tightness around the $0$-value.
    We demonstrate in section \ref{sec:model_overview} how it is incorporated in model training despite the fact that $u^{(n_v)}$ above is latent.
\ei
We finally note that the number/type of (virtual) observations can differ for each data-pair \cite{zang2025dgenno}.

\subsection{Deep generative neural operator for the SP linkage}
\label{sec:model_dgno}
\subsubsection{Overview of the proposed probabilistic framework}
\label{sec:model_overview}
\begin{figure}[tb]
\centering
\includegraphics[width=0.95\textwidth]{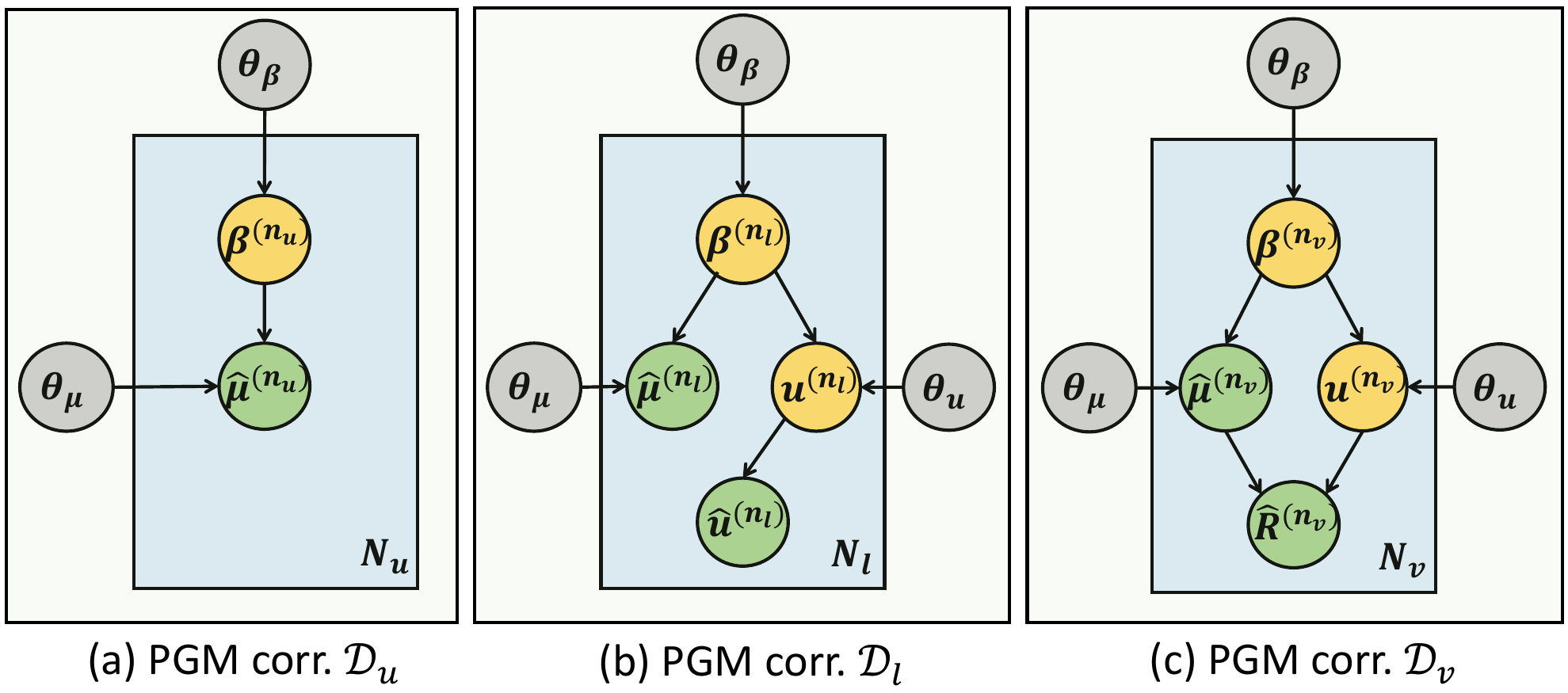}
\caption{Probabilistic graphical model (PGM) illustration for the proposed framework. The nodes in yellow represent latent/unobserved variables, the nodes in green correspond to observed variables, and the nodes in gray represent learnable parameters. The arrows indicate dependencies, with the parent node conditioning the child node. The conditional densities defined within the model are described in the main text. Together, they encode the forward mapping and are trained as described in Section \ref{sec:train}.}
\label{fig:Prop_Graph}
\end{figure}

An overview of the probabilistic graphical model for the proposed framework is shown in Figure \ref{fig:Prop_Graph}. At the core of the formulation is a finite-dimensional latent variable vector
$\bm{\beta} \in \mathbb{R}^{d_{\beta}}$,
which acts as a common generator for both the PDE input (the microstructure $\mu$) and the PDE solution (the solution field $u$). 
Operating in this lower-dimensional, well-structured latent space, $\bm{\beta}$ replaces the high-dimensional microstructure $\mu$ both
in the solution of forward and inverse problems. 

We first describe the main building blocks of the proposed model before discussing the resulting likelihoods of the data modalities that are used for training. For additional details, we refer readers to \cite{zang2025dgenno}.
In brief, the model relies on three densities, namely:
\bi
\item $p_{\bt_{\bmb}}(\bmb)$ which serves as the learnable prior on the latent parameters $\bmb$ which is parametrized by $\bt_{\bmb}$. Details are provided in section \ref{sec:prior}.
\item $p_{\bt_{\mu}}(\mu|\bmb)$ which serves as a learnable decoder for the PDE-inputs, i.e., the microstructures $\mu$ given the latent variables $\bmb$, which is parametrized by $\bt_{\mu}$,
\item $p_{\bt_{u}}(u|\bmb)$ which serves as a learnable decoder for the PDE-outputs, i.e. $u$ given the latent variables $\bmb$, which is parametrized by $\bt_{u}$.
\ei

\begin{figure}[tb]
\centering
\includegraphics[width=1.\textwidth]{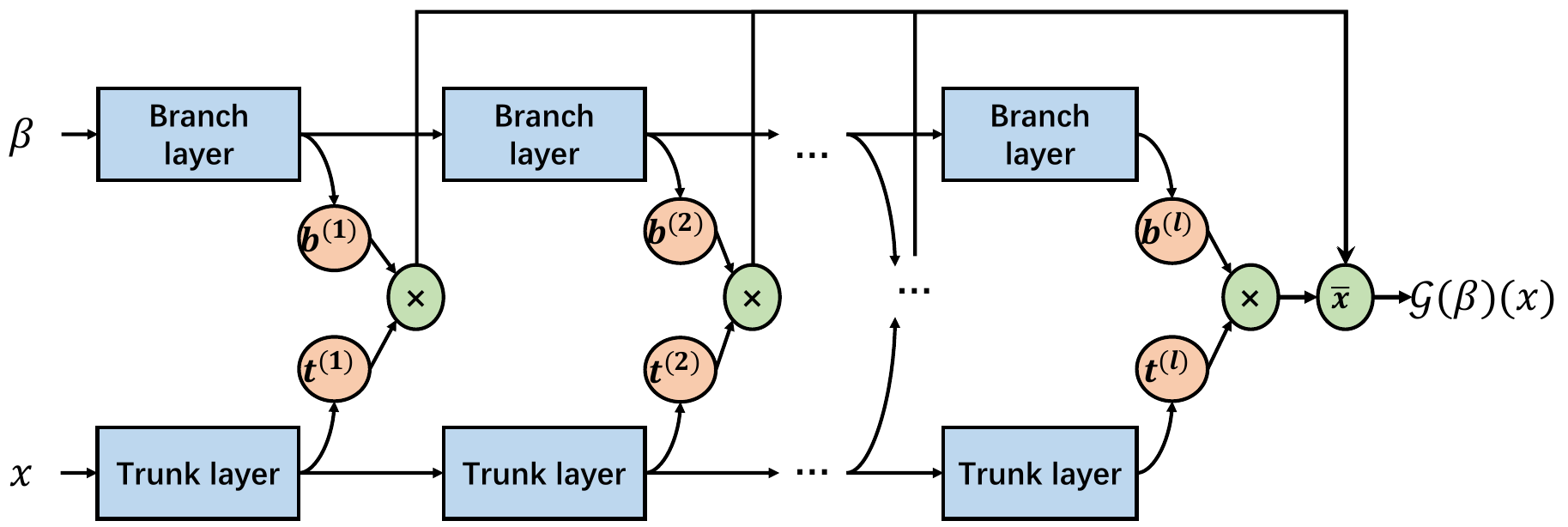}
\caption{The MultiONet architecture.}
\label{fig:MultiONet}
\end{figure}

The decoders $p_{\bm{\theta}_\mu}(\mu|\bm{\beta})$ and $p_{\bm{\theta}_u}(u|\bm{\beta})$ map the latent representation $\bm{\beta}$ to the PDE-input field $\mu$ and to the PDE-output field $u$. 
Since learning these mappings involves translating a finite-dimensional latent vector into infinite-dimensional outputs, standard architectures such as MLPs or CNNs are insufficient. To address this, we have introduced  the MultiONet architecture~\cite{zang2025dgenno}, an extension of DeepONet \cite{lu_learning_2021}, designed for efficient and expressive operator learning. A schematic of the MultiONet is shown in Figure~\ref{fig:MultiONet}, with its mathematical formulation detailed in Section~\ref{sec:model_for_multionet}.
As in \cite{zang2025dgenno}, we introduce two such MuliONet neural operators $\mathcal{G}_{\bm{\theta}_u}$ and $\mathcal{G}_{\bm{\theta}_\mu}$ which are  parameterized by $\bm{\theta}_u$ and $\bm{\theta}_\mu$ respectively. They are used to model the decoders above as:
\begin{equation}
p_{\bm{\theta}_u}({u}(\bx)|\bm{\beta})
= \delta\!\left( {u}(\bm{x}) - \mathcal{G}_{\bm{\theta}_u}(\bm{\beta})(\bm{x}) \right),
\label{eq:decodeu}
\end{equation}
Furthermore, for a two-phase medium, i.e., a binary-valued field 
\(\mu\) with \(\mu(\bx) \in \{0,1\}\) for all \(\bx \in \Omega\), as used in our numerical examples—and for discrete grid locations \(\{\bm{x}_j\}_{j=1}^{J}\), we model the conditional distribution as
\begin{equation}\label{eq:mu_pred}
p_{\bm{\theta}_\mu}(\mu \mid \bm{\beta})
= \prod_{j=1}^J p\big(\mu(\bm{x}_j)\mid \bm{\beta}\big),
\qquad
p\big(\mu(\bm{x}_j)=1\mid \bm{\beta}\big) = \sigma\!\big( \mathcal{G}_{\bm{\theta}_\mu}(\bm{\beta})(\bm{x}_j)\big),
\end{equation}
where \(\sigma(\cdot)\) is the logistic sigmoid. That is, \(\mathcal{G}_{\bm{\theta}_\mu}\) outputs the logits for each grid location, and the microstructure is treated as a collection of independent Bernoulli random variables.
 Further details and extensions to multi-phase media can be found in \cite{zang2025dgenno}.

We denote collectively with $\bt=\{\bt_{\bmb},\bt_{\mu}, \bt_{u}\}$ the model parameters associated with  three densities above. We present the likelihood of each of the three data modalities described in section \ref{sec:training_data}. In particular:
\bi
\item For the unlabeled data $\mathcal{D}_u=\{ \hat{\mu}^{(n_u)} \}_{n_u=1}^{N_u}$:
\be
p_{\bt}(\{ \hat{\mu}^{(n_u)} \}_{n_u=1}^{N_u}\})=\prod_{n_u=1}^{N_u} \int p_{\bt_{\mu}} (\hat{\mu}^{(n_u)} | \bm{\beta}^{(n_u)})~p_{\bt_{\bmb}} (\bmb^{(n_u)})~d\bmb^{(n_u)}.
\label{eq:likeu}
\ee
\item For the labeled data pairs  $\mathcal{D}_l=\{ \hat{\mu}^{(n_l)}, \hat{\bm{u}}^{(n_l)} \}_{n_l=1}^{N_l}$:
\begin{align}
p_{\bt}(\{ \hat{\mu}^{(n_l)}, \hat{\bm{u}}^{(n_l)} \}_{n_l=1}^{N_l}) 
& =\prod_{n_l=1}^{N_l} \int p(\hat{\bm{u}}^{(n_l)} |  u^{(n_l)}) p_{\bt_u}(u^{(n_l)}| \bm{\beta}^{(n_l)}) ~ p_{\bt_{\mu}} (\hat{\mu}^{(n_l)} | \bm{\beta}^{(n_l)} ) \notag \\
&\quad \times p_{\bt_{\bmb}} (\bmb^{(n_l)})~du^{(n_l)}~d\bmb^{(n_l)} \\
& =\prod_{n_l=1}^{N_l} \int p(\hat{\bm{u}}^{(n_l)} |  \mathcal{G}_{\bm{\theta}_u}(\bm{\beta}^{(n_l)})~ p_{\bt_{\mu}} (\hat{\mu}^{(n_l)} | \bm{\beta}^{(n_l)} ) \notag \\
&\quad \times p_{\bt_{\bmb}} (\bmb^{(n_l)})~d\bmb^{(n_l)}
\label{eq:likel}
\end{align}
where the decoder in \refeqp{eq:decodeu} is used, as well as the likelihood for labeled data is given in \refeqp{eq:ulike}.
\item For the virtual data pairs $\mathcal{D}_v=\{ \hat{\mu}^{(n_v)}, \hat{\bs{R}}^{(n_v)} \}_{n_v=1}^{N_v}$:
\begin{align}
p_{\bt}\!\left(\{ \hat{\mu}^{(n_v)}, \hat{\bs{R}}^{(n_v)} \}_{n_v=1}^{N_v}\right)
&= \prod_{n_v=1}^{N_v} \int 
p(\hat{\bs{R}}^{(n_v)} \mid u^{(n_v)},\hat{\mu}^{(n_v)} ) p_{\bt_u}(u^{(n_v)}| \bm{\beta}^{(n_v)}) \notag \\
&\quad \times p_{\bt_{\mu}} (\hat{\mu}^{(n_v)} | \bm{\beta}^{(n_v)} )
p_{\bt_{\bmb}}\!(\bmb^{(n_v)})\, ~du^{(n_v)} d\bmb^{(n_v)} \\
&= \prod_{n_v=1}^{N_v} \int 
p\!\left(\hat{\bs{R}}^{(n_v)} \mid \mathcal{G}_{\bm{\theta}_u}(\bm{\beta}^{(n_v)}), \hat{\mu}^{(n_v)}\right) \notag \\
&\quad \times p_{\bt_{\mu}} (\hat{\mu}^{(n_v)} | \bm{\beta}^{(n_v)} )\,
p_{\bt_{\bmb}}\!(\bmb^{(n_v)})\,d\bmb^{(n_v)}
\label{eq:likev}
\end{align}
where the decoder in \refeqp{eq:decodeu} is used as well as  the  {\em virtual} likelihood  in \refeqp{eq:vlike}.
\ei
A general observation that is of relevance during training is that $\bmb$'s are latent and would need to be inferred.

\subsubsection{Latent variables and prior}
\label{sec:prior}
The latent variables $\bm{\beta}$ play a central role in the proposed framework, providing a low-dimensional and well-structured representation of both the original high-dimensional and irregular microstructure $\mu$ and the corresponding solution field $u$. Classical approaches for encoding microstructures rely mainly on predefined descriptors, such as texture statistics \cite{koutsourelakis_probabilistic_2006,haralick2007textural}, grain size distribution \cite{backman2006icme}, auto- and cross-correlation functions \cite{kalidindi2011microstructure}, or dimensionality-reduction techniques such as PCA \cite{xu2017novel}. However, these descriptors are typically agnostic to the PDE solution $u$—and thus to the target property of interest—because they only capture geometric or statistical information from microstructure $\mu$ without directly incorporating property-related correlations.

The prior distribution $p_{\bm{\theta}_\beta}(\bm{\beta})$ is often fixed as a standard Gaussian or uniform density in conventional latent-variable models. However, such assumptions impose strong constraints on the geometry of the latent space and may fail to capture the true, potentially multi-modal distribution of $\bm{\beta}$ induced by the data \cite{wang_uncertainty-aware_2025}. To overcome this limitation, we adopt a flexible, learnable prior parameterized by a normalizing flow (NF)~\cite{tabak_family_2012,rezende_variational_2015}. NFs model complex densities via a sequence of invertible and computationally tractable transformations, allowing the prior to adapt to the underlying data distribution. This adaptive prior leads to a more expressive latent representation and improves the accuracy and diversity of the generative modeling process.

Specifically, we use the Real-valued Non-Volume Preserving (RealNVP) flow \cite{dinh2016density}, denoted by $f_{\bmtheta_\beta}$ and parametrized by $\bmtheta_\beta$, which consists of a stack of invertible affine coupling layers. By alternating the partitioning across layers, RealNVP achieves high expressiveness while retaining efficient invertibility.
We train the RealNVP model $f$ such that it maps the latent variables $\bm{\beta}$ to a standard multivariate normal distribution,
\begin{equation}
f_{\bmtheta_\beta}(\bm{\beta})=\bm{z} \sim \mathcal{N}(\mathbf{0}, \mathbf{I}).
\end{equation}
Once trained, the prior distribution $p_{\bm{\theta}_\beta}(\bm{\beta})$ is implicitly defined as the inverse mapping of a Gaussian sample:
\begin{equation}\label{eq:NF}
\bm{\beta} = f_{\bmtheta_\beta}^{-1}(\bm{z}), \quad \bm{z} \sim \mathcal{N}(\mathbf{0}, \mathbf{I}).
\end{equation}
This formulation enables us to draw samples from a data-adaptive prior while retaining the computational simplicity of Gaussian sampling in the $z$-space, ultimately leading to a more realistic representation of microstructure variability and improved predictive accuracy of our model.

\subsection{Model training}
\label{sec:train}
Training of the model, i.e., identifying the optimal values for the model parameters $\bt$, is carried out by maximum likelihood. The total log-likelihood arises from the contributions of the three terms in Equations \eqref{eq:likeu}, \eqref{eq:likel}, and \eqref{eq:likev}. 
We note that all three likelihood terms involve intractable integrations with respect to the latent variables $\bmb$. To this end, we employ Stochastic Variational Inference (SVI, \cite{paisley_variational_2012}), which involves maximizing lower-bounds to the log-likelihood terms that arise from the introduction of an auxiliary density for the latent variables and by Jensen's inequality \cite{bishop_pattern_2007}.  In particular, we adopt an amortized version of SVI with an auxiliary, degenerate density $q_{\bm{\eta}}$:
\begin{equation}
q_{\bm{\eta}}(\bmb | \hat{\mu}) =
\delta \big( \bm{\beta} - e_{\bm{\eta}}(\hat{\mu}) \big),
\label{eq:q}
\end{equation}
where $e_{\bm{\eta}}(\hat{\mu})$ is an encoder network parametrized by $\bm{\eta}$ \cite{zang2025dgenno}. With this in hand, the lower-bounds of each of the likelihood terms become\footnote{The notation $<.>_q$ suggests expectations with respect to $q$.}:
\bi
\item For the unlabeled data $\mathcal{D}_u=\{ \hat{\mu}^{(n_u)} \}_{n_u=1}^{N_u}$:
\be
\begin{array}{ll}
\log p_{\bt}(\{ \hat{\mu}^{(n_u)} \}_{n_u=1}^{N_u}\}) & =\sum_{n_u=1}^{N_u}  log \int p_{\bt_{\mu}} (\hat{\mu}^{(n_u)} | \bm{\beta}^{(n_u)})~p_{\bt_{\bmb}} (\bmb^{(n_u)})~d\bmb^{(n_u)} \\
& \ge \sum_{n_u=1}^{N_u}   \left<  \log \cfrac{p_{\bt_{\mu}} (\hat{\mu}^{(n_u)} | \bm{\beta}^{(n_u)})~p_{\bt_{\bmb}} (\bmb^{(n_u)})}{q_{\bm{\eta}}(\bmb^{(n_u)} | \hat{\mu}^{(n_u)})   } \right>_{q_{\bm{\eta}}(\bmb^{(n_u)} | \hat{\mu}^{(n_u)})} \\
& = \mathcal{L}_u(\bt,\bm{\eta}).
\end{array}
\label{eq:likeuelbo}
\ee
\item For the labeled data pairs  $\mathcal{D}_l=\{ \hat{\mu}^{(n_l)}, \hat{\bm{u}}^{(n_l)} \}_{n_l=1}^{N_l}$:
\be
\begin{array}{ll}
\log p_{\bt}(\{ \hat{\mu}^{(n_l)}, \hat{\bm{u}}^{(n_l)} \}_{n_l=1}^{N_l}) & =\sum_{n_l=1}^{N_l}\left( \log \int p(\hat{\bm{u}}^{(n_l)} |  \mathcal{G}_{\bm{\theta}_u}(\bm{\beta}^{(n_l)}))~ p_{\bt_{\mu}} (\hat{\mu}^{(n_l)} | \bm{\beta}^{(n_l)} )\right. 
\\
& \qquad \qquad  \qquad  \times  \left. p_{\bt_{\bmb}} (\bmb^{(n_l)})~ d\bmb^{(n_l)} \right) \\
& \geq \sum_{n_l=1}^{N_l} \left( \left< \log  p(\hat{\bm{u}}^{(n_l)} |  \mathcal{G}_{\bm{\theta}_u}(\bm{\beta}^{(n_l)}))~ p_{\bt_{\mu}} (\hat{\mu}^{(n_l)} | \bm{\beta}^{(n_l)} )  \right>_{q_{\bm{\eta}}(\bmb^{(n_l)} | \hat{\mu}^{(n_l)})} \right. \\
&  \qquad  \qquad \left. \left< \log \cfrac{~p_{\bt_{\bmb}} (\bmb^{(n_l)})}{q_{\bm{\eta}}(\bmb^{(n_l)} | \hat{\mu}^{(n_l)})} \right>_{q_{\bm{\eta}}(\bmb^{(n_l)} | \hat{\mu}^{(n_l)})}  \right) \\
& = \mathcal{L}_l(\bt, \bm{\eta}).
\end{array}
\label{eq:likelelbo}
\ee
\item For the virtual data pairs $\mathcal{D}_v=\{ \hat{\mu}^{(n_v)}, \hat{\bs{R}}^{(n_v)} \}_{n_v=1}^{N_v}$:
\be
\begin{array}{ll}
\log p_{\bt}\!\left(\{ \hat{\mu}^{(n_v)}, \hat{\bs{R}}^{(n_v)} \}_{n_v=1}^{N_v}\right)
&= \sum_{n_v=1}^{N_v} \log \int 
p\!\left(\hat{\bs{R}}^{(n_v)} \mid \mathcal{G}_{\bt_u}(\bmb^{(n_v)}), \hat{\mu}^{(n_v)}\right)\, \\
&\quad\quad\quad\quad\quad\quad \times p_{\bt_{\mu}}\left(\hat{\mu}^{(n_v)} \mid \bmb^{(n_v)}\right)\,
p_{\bt_{\bmb}}\!\left(\bmb^{(n_v)}\right)\,d\bmb^{(n_v)} \\
& \ge \sum_{n_v=1}^{N_v} \left( \left< \log p\!\left(\hat{\bs{R}}^{(n_v)} \mid \mathcal{G}_{\bm{\theta}_u}(\bm{\beta}^{(n_v)}) , \hat{\mu}^{(n_v)}\right)\,
\right>_{q_{\bm{\eta}}(\bmb^{(n_v)} | \hat{\mu}^{(n_v)})} \right. \\
&\quad \left. + \left< \log  \cfrac{p_{\bt_{\mu}}\left(\hat{\mu}^{(n_v)} \mid \bmb^{(n_v)}\right)\,
p_{\bt_{\bmb}}\!\left(\bmb^{(n_v)}\right)\ }{q_{\bm{\eta}}(\bmb^{(n_v)} | \hat{\mu}^{(n_v)}) }  \right>_{q_{\bm{\eta}}(\bmb^{(n_v)} | \hat{\mu}^{(n_v)}) } \right) \\
& = \mathcal{L}_{v}(\bt,\bm{\eta}).
\end{array}
\label{eq:likevelbo}
\ee
\ei

The combination of the aforementioned lower-bounds leads to the model objective $\mathcal{L}$, i.e.:
\be\label{eq:loss}
\mathcal{L}(\bt,\bm{\eta})=\mathcal{L}_u(\bt, \bm{\eta})+\mathcal{L}_l(\bt, \bm{\eta})+\mathcal{L}_{v}(\bt,\bm{\eta}),
\ee
which accounts for all data modalities.
Obviously, if any of the data modalities are absent (as, e.g., $\mathcal{D}_l$, in the numerical illustration of section \ref{sec:maxdesign}), the corresponding term ($\mathcal{L}_l$ in section \ref{sec:maxdesign}) would not need to be included.
Detailed expressions of the aforementioned terms in $\mathcal{L}$  for the specific inverse design problems considered are provided in \ref{sec:loss_terms}.
In order to maximize $\mathcal{L}$, we employ a stochastic gradient-ascent scheme, e.g., ADAM, and subsample mini-batches of the data as described in Algorithm~\ref{alg:foward} where the overall training procedure is summarized. The implementation of Design-GenNO will be released in a public GitHub repository at \href{https://github.com/yaohua32/Design-GenNO}{https://github.com/yaohua32/Design-GenNO} upon acceptance of this paper.
\begin{algorithm}[!t]
\caption{The training of the proposed model}\label{alg:foward}
\begin{algorithmic}
\Inputs{Training dataset $\mathcal{D}=\mathcal{D}_u\cup\mathcal{D}_l\cup\mathcal{D}_v$; Weights $\lambda_{pde}, \lambda_{data}, \lambda_{kl}$}
\Initialize{Model parameters $\bmtheta=(\bmtheta_\mu, \bmtheta_u, \bmtheta_\beta)$ and $\bm{\eta}$, learning rate $lr$.}
\While{Convergence or maximum number of iterations not reached}
\For{each mini-batch $\mathcal{S} \subset \mathcal{D}$}
\State {Encode microstructures: $\bm{\beta}=e_{\bm{\eta}}(\hat{\mu}), \quad \hat{\mu} \in \mathcal{S}$} \hfill \refeqp{eq:q}
\State {Predict field responses: $u=\mathcal{G}_{\bt_u}(\bm{\beta})$} \hfill \refeqp{eq:decodeu}
\State {Reconstruct microstructures: $\mu=\mathcal{G}_{\bt_\mu}(\bm{\beta})$} \hfill \refeqp{eq:mu_pred}
\State {Compute weighted PDE residuals (Eq.~\eqref{eq:weak_residual}): $\hat{r}_{m}, \ m=1,\dots,M$}
\State {Evaluate mini-batch ELBO (\refeqp{eq:loss}): $$\mathcal{L}=\mathcal{L}_{u}+\mathcal{L}_{l}+\mathcal{L}_{v}$$} 
\State {Update parameter $\bm{
\eta
}$, $\bmtheta$ with stochastic gradient ascent:
$$\bm{\eta} \leftarrow \bm{\eta} + lr \odot \nabla_{\bm{\eta}} \mathcal{L}(\bmtheta,\bm{\eta}).$$
$$\bmtheta \leftarrow \bmtheta + lr \odot \nabla_{\bmtheta} \mathcal{L}(\bmtheta,\bm{\eta}).$$ }
\EndFor
\If{at every 200th epoch}
\State {Update the learning rate to $lr \leftarrow lr/2$.}
\EndIf
\EndWhile
\end{algorithmic}
\end{algorithm}

\subsection{Solving the inverse design problem}
\label{sec:inverse_design}
Once the model has been trained and the optimal values for the parameters $\bm{\theta}^*$ and $\bm{\eta}^*$ of the generative and inference model have been obtained, the resulting framework can serve as a fully differentiable surrogate that can provide probabilistic predictions to both forward and inverse problems {\em without} the need of additional model simulations. The key lies in the latent variables $\bmb$, which serve as common generators and disentangle the SP relation between the microstructure $\mu$ and the corresponding response $u$.
Consequently, all inverse design problems, such as  \textbf{P1-P3} introduced in section \ref{sec:defp13}, can be solved using a unified two-step procedure consistent with the formulation of a Bayesian inverse problem \cite{ikebata_bayesian_2017}. 
In the first step, inference is carried out with respect to $\bmb$ and from a posterior density that depends on the particulars of the design problem. Whether one is interested in point estimates or in drawing samples of $\bmb$, the availability of derivatives with respect to $\bmb$ provides a major advantage, allowing the associated tasks to be executed efficiently.
In the second step, which is common to {\em all} inverse design problems considered, the inferred samples of $\bm{\beta}$ are propagated through the trained decoder $p_{\bm{\theta}_\mu^*}(\mu|\bmb)$ of \refeqp{eq:mu_pred} to generate microstructure samples $\mu$, which are consistent with the design objectives.
We provide the details and justifications for each of the three design problems of interest below.
\subsubsection{Solution for the problem \textbf{(P1)}}
\label{sec:sol_p1}
We first consider the problem \textbf{(P1)} where the goal is to obtain samples from $p(\mu | \ked)$, i.e., microstructures whose properties lie in the target domain $\mathcal{T}_d$.
According to the trained model,  the joint posterior $p_{\bt^*}(\mu, \bmb | \ked)$ can be written as:
\be\label{eq:p1_joint}
\begin{array}{ll}
p_{\bt^*}(\mu, \bmb | \ked) & \propto p(\ked|\bmb, \mu ) p_{\bt^*}(\mu , \bmb) \\
&=  \left( \int p(\ked, u| \bmb, \mu) ~du \right)~~p_{\bt^*_\mu}(\mu | \bmb) p_{\bt^*_\beta}(\bmb) \\
&= \left( \int p(\ked | u) p_{\bt^*_u}(u|\bmb)~du \right)~~p_{\bt^*_\mu}(\mu | \bmb) p_{\bt^*_\beta}(\bmb) \\
&= \left( \int p(\ked | u) p_{\bt^*_u}(u| \bmb) ~du~p_{\bt^*_\beta}(\bmb) \right)~p_{\bt^*_\mu}(\mu | \bmb)  \\
& = p(\bmb | \ked)~~p_{\bt^*_\mu}(\mu | \bmb).
\end{array}
\ee
Here we have made use of the crucial model property that given $\bmb$, $\mu$, and $u$ are conditionally independent (line 3) and therefore decouple in the ensuing expressions.  Hence one simply needs first to sample/infer $\bmb$ from the posterior $p(\bmb | \ked)$ (with derivatives - see below) and then readily propagate forward through the decoder $p_{\bm{\theta}_\mu^*}(\mu | \bmb)$ to get as many samples of $\mu$ as needed.

With regards to the posterior $p(\bmb | \ked) \propto p (\ked|\bmb) p_{\bt^*_\beta}(\bm{\beta})$, if $\kappa_{eff}(u)$ is the known function that gives properties and given the decoder $p_{\bt^*_u}(u|\bmb)$ in \refeqp{eq:decodeu}, the likelihood $p(\ked|\bmb)$ will be:
\be\label{eq:likekeq}
p(\ked|\bmb)=\int p(\ked | u ) p_{\bt^*_u}(u| \bmb) ~du = 
\begin{cases}
1, \quad \textrm{if $\kappa_{eff}\left( \mathcal{G}_{\bm{\theta}^*_u}(\bm{\beta}) \right) \in \mathcal{T}_d$} \\
0, \quad \textrm{~~otherwise}
\end{cases}
\ee
This formulation is differentiable but not informative, as the derivatives with respect to $\bmb$ are always zero. While methods such as MCMC or variational inference (VI) can still be employed to infer $\bmb$, the lack of useful gradient information can make the process inefficient, particularly when $\mathrm{dim}(\bmb) \gg 1$. This issue becomes especially critical for $\bmb$ samples that do not yield $\kappa_{\mathrm{eff}} \in \mathcal{T}_d$, as, without informative gradients, it becomes difficult to effectively guide these samples toward the target domain.
To this end, we employ a commonly used mollification technique~\cite{koutsourelakis_design_2008}, in which we approximate the likelihood as:
\begin{equation}
p(\kappa_{\mathrm{eff}}|\bmb) =
\begin{cases}
1, & \text{if } \kappa_{\mathrm{eff}}\left( \mathcal{G}_{\bm{\theta}^*_u}(\bmb) \right) \in \mathcal{T}_d, \\
\exp\!\left(-\tau \, d(\bmb)\right), & \text{otherwise},
\end{cases}
\end{equation}
where $d(\bmb)$ denotes a differentiable distance from the target domain $\mathcal{T}_d$, such that $d(\bmb) = 0$ when the corresponding properties $\kappa_{eff}\left( \mathcal{G}_{\bm{\theta}^*_u}(\bm{\beta}) \right)$  lie within $\mathcal{T}_d$ and $d(\bmb) \to +\infty$ as the distance from $\mathcal{T}_d$ increases. The tempering parameter $\tau>0$ controls the strength of the penalty: as $\tau \to  +\infty$, the expression recovers the hard constraint in Eq.~\eqref{eq:likekeq}, whereas smaller values of $\tau$ produce a smoother, more gradual transition. Importantly, for any finite $\tau$, the likelihood remains differentiable with respect to $\bmb$, enabling gradient-based updates that effectively push the samples toward the target domain.   

\subsubsection{Solution for the problem \textbf{(P2)}}
\label{sec:sol_p2}
In the second problem \textbf{(P2)}, the objective is to generate microstructures that result in a desired response field ${u}_d$. According to the trained DGenNO model,  the joint posterior $p_{\bt^*}(\mu, \bmb | {u}_d)$ can be written as:
\be
\begin{array}{ll}
p_{\bt^*}(\mu, \bmb | {u}_d) & \propto p( {u}_d |\mu, \bmb ) p_{\bt^*}(\mu, \bmb ) \\
& = \left( \int p( {u}_d, u |\mu, \bmb ) ~du\right) ~~p_{\bt^*_\mu}(\mu |\bmb ) p_{\bt^*_\beta}(\bmb) \\ 
& = \left( \int p( {u}_d | u) p_{\bt^*_u}(u  |\bmb ) ~du\right) ~~p_{\bt^*_\mu}(\mu |\bmb ) p_{\bt^*_\beta}(\bmb) \\  
& =  \left( \int p( {u}_d | u) p_{\bt^*_u}(u | \bmb ) ~du~ p_{\bt^*_\beta}(\bmb)\right) ~~p_{\bt^*_\mu}(\mu |\bmb ) \\
& = p(\bmb | {u}_d) ~~p_{\bt^*_\mu}(\mu |\bmb ).
\end{array}
\ee
Here we have again made use of the conditional independence of $\mu$ and $u$ given $\bmb$ (line 3) in order to arrive to the final expression that requires sampling $\bmb$ from the corresponding posterior $p(\bmb | {u}_d) \propto  \int p( {u}_d | u) p_{\bt^*}(u | \bmb ) ~du~ p_{\bt^*_\beta}(\bmb)$. Assuming that 
$p({u}_d | u) \propto \exp\{-\tau_u ||{u}_d-o(u)||^2\} $ where $o(.)$ denotes the observation filter and $\tau_u^{-1}$ determines the level of proximity to ${u}_d$ desired, we obtain that: 
\be
p(\bmb | {u}_d) \propto e^{-\tau_u \|{u}_d-o(\mathcal{G}_{\bm{\theta}^*_u}(\bm{\beta}))\|^2} ~p_{\bt^*_\beta}(\bmb)
\ee
Hence, in the first step, we perform inference on $\bmb$ based on this posterior, which can be readily carried out thanks to its differentiability—a direct consequence of the differentiable neural operator $\mathcal{G}_{\bm{\theta}^*_u}(\bmb)$. In the second step, as in previous cases, the sampled $\bmb$  are propagated through the decoder $p_{\bt^*_\mu}(\mu | \bmb)$ to generate microstructures consistent with the target response.
\subsubsection{Solution for the problem \textbf{(P3)}}
\label{sec:sol_p3}
Similar to problems \textbf{(P1)} and \textbf{(P2)}, and under the adopted probabilistic formulation, the solution of the maximization problem \textbf{(P3)} can be cast as a Bayesian inverse problem. To this end, and given that the sought maximum of $F_{\kappa_{eff}}:=F(\kappa_{eff})>0$ is a priori unknown, we employ a likelihood of the form $p(F_{\kappa_{eff}}|u) = F_{\kappa_{eff}}^{\alpha}(u)$ from \cite{bissiri2016general}. We note that $\alpha$ serves as a tempering parameter, and as $\alpha \to \infty$, the likelihood is more tightly concentrated around the (global) maximum of $F_{\kappa_{eff}}(u)$. With some abuse of terminology and notation, we denote with $p_{\bt^*}(\bmb, \mu | F_{\kappa_{eff}})$ the corresponding joint (posterior) density on $\bmb, \mu$, which, based on the likelihood adopted, can be written as:
\be\label{eq:p3_joint}
\begin{array}{ll}
p_{\bt^*}(\bmb, \mu | F_{\kappa_{eff}}) & \propto p ( F_{\kappa_{eff}} |\bmb, \mu ) ~p_{\bt^*}(\bmb, \mu ) \\
& = \left( \int p ( F_{\kappa_{eff}}, u |\bmb , \mu)~du \right)~ ~p_{\bt^*_\mu}(\mu|\bmb) p_{\bt^*_\beta}(\bmb)\\
& = \left( \int p ( F_{\kappa_{eff}} | u ) p_{\bt^*_u}(u | \bmb )~du \right)~ ~p_{\bt^*_\mu}(\mu|\bmb) p_{\bt^*_\beta}(\bmb) \\
& =  \left( \int F_{\kappa_{eff}}^{\alpha}(u) p_{\bt^*_u}(u|\bmb) ~du ~p_{\bt^*_\beta}(\bmb) \right)~~p_{\bt^*_\mu}(\mu|\bmb) \\
& = \left( F_{\kappa_{eff}}^{\alpha}\left(\mathcal{G}_{\bm{\theta}^*_u}(\bm{\beta}) \right)~p_{\bt^*_\beta}(\bmb) \right)~~p_{\bt^*_\mu}(\mu |\bmb) \\
 & \propto  p( \bmb | F_{\kappa_{eff}}) ~~~p_{\bt^*_\mu}(\mu|\bmb)
\end{array}
\ee
where we have made use of the learned decoder $p_{\bt^*_u}(u|\bmb)$ in \refeqp{eq:decodeu}. More importantly, and as in the previous cases, this formulation enables the decomposition of the solution into two tractable steps:
\bi
\item Sample $\bmb$ from the unnormalized (posterior) density given by $p( \bmb | F_{\kappa_{eff}}) \propto F_{\kappa_{eff}}^{\alpha}\left(\mathcal{G}_{\bm{\theta}^*_u}(\bm{\beta}) \right)~p_{\bt^*_\beta}(\bmb)$. All terms are differentiable and $\bmb$ is continuous-valued, ensuring the availability of derivatives and thereby facilitating faster inference.
\item Propagate those samples through the learned decoder $p_{\bt^*_\mu}(\mu|\bmb)$ (\refeqp{eq:mu_pred}) in order to obtain corresponding microstructures.
\ei
In this paper, we employed $\alpha=10$, which was found empirically to be sufficient.  We note finally that in cases where the property objective exhibits multiple local minima, various tempering schemes with respect to $\alpha$ can be employed to identify all such regions in the first step above.

\section{Experiments}
\label{sec:experiments}
In this section, we assess the effectiveness of the proposed Design-GenNO for inverse microstructure design in two-phase thermally conductive materials, considering diverse design targets such as effective thermal conductivity and temperature distribution. The microstructure generation process and the SP linkage are described in Section \ref{sec:experiment_sp}. As noted in Section \ref{sec:introduction}, existing PIML approaches face difficulties in microstructure-centered design, especially when targeting effective properties, while current DNO-based methods rely solely on data-driven training and are thus either not directly comparable or incapable of addressing the design problems considered here. To provide a meaningful baseline, we compare our approach against \textbf{PoreFlow}\cite{mirzaee2025inverse}, a recently proposed,  state-of-the-art method specifically designed for microstructure inverse design with effective property targets.
{PoreFlow} is a conditional generative model that uses a conditional normalizing flow (CNF) to learn the distribution of latent representations $\bm{\beta}$ corresponding to high-dimensional microstructures $\mu$. The model is trained in a purely data-driven manner by minimizing a composite loss comprising: (i) a reconstruction loss for recovering the microstructure from $\bm{\beta}$, (ii) a regression loss for predicting the target property $\kappa_{eff}$, and (iii) a KL divergence loss, analogous to \eqref{eq:loss_kl}, for regularizing the NF model. Once trained, latent vectors $z \in \mathbb{R}^{d_\beta}$ sampled from a Gaussian prior are transformed by the CNF, conditioned on the target effective properties, into target-specific latent variables, which are then decoded into microstructures. Despite its purely data-driven training, PoreFlow is chosen as a baseline because it tackles a similar microstructure inverse design problem with target effective properties and leverages deep generative models for compact, meaningful design representations. Details of model setups for both methods and the generation of training data are provided in Section~\ref{sec:gen_data}.
\subsection{Experimental setup}
\label{sec:setup}
\subsubsection{The forward simulation}
\label{sec:experiment_sp}
The microstructure $\mu$ is represented as a $32 \times 32$ binary image, with the thermal conductivities of phase 1 and phase 2 set to $\kappa_1 = 10$ and $\kappa_2 = 2$, respectively. Microstructures are generated using a processing–structure (PS) simulation framework \cite{zang2025dgenno} involving a Gaussian Random Field (GRF) and a phase-field model \cite{zang2025psp}. First, a random image is sampled from a GRF with the spectral density function:
$$
S(\phi) = \phi e^{-k_x^2} + (10 - \phi) e^{-k_y^2}, \quad \phi \in [1, 9],
$$
where $k_x, k_y$ represent shifted Fourier coordinates. Then, it was followed by thresholding to obtain a binary image with the volume fraction of phase 1 being $1/3$. This binary mask is then evolved using a Cahn–Hilliard equation:
$$
\frac{\partial c}{\partial t} = D \nabla^2 \left( c^3 - c - \gamma \nabla^2 c \right), \quad t \in [0, 10],
$$
where $c = \pm 1$ indicates the concentration of the two phases, $D = 50$ is the diffusion coefficient, and $\gamma= 1$ controls the interface thickness. This procedure can efficiently generate large numbers of physically consistent microstructures at negligible cost.

The SP linkage is governed by the following elliptic PDE:
\begin{equation} \label{eq:pde}
\nabla \cdot \left( \mu(\bm{x}) \nabla T(\bm{x}) \right) = 0, \quad \bm{x} \in \Omega=[0,1]^2,
\end{equation}
subject to mixed boundary conditions:
\begin{equation}\label{eq:pde_bd}
T(\bm{x}) = T_D(\bm{x}), \quad \bm{x} \in \partial \Omega_D; \quad
\mu(\bm{x}) \nabla T(\bm{x}) \cdot \bm{n} = 0, \quad \bm{x} \in \partial \Omega_N,
\end{equation}
where $T$ represents the temperature field, $\partial \Omega_D$ and $\partial \Omega_N$ denote the Dirichlet and Neumann boundaries, respectively, and $\bm{n}$ is the unit outward normal vector.

For problems \textbf{(P1)} and \textbf{(P3)}, the forward map is used to compute the effective conductivity $\kappa_{eff}=(\kappa_h,\kappa_v)$, where $\kappa_h$ and $\kappa_v$ represent the horizontal and vertical conductivities. For example, to compute $\kappa_h$, a horizontal temperature gradient is imposed by setting $T = 0$ on the left boundary and $T = 1$ on the right. The resulting solution $T(\bm{x})$ is then used to compute $\kappa_h$:
\begin{equation} \label{eq:kappa_h}
\kappa_h = -\frac{1}{|\Omega|} \int_{\Omega} \mu(\bm{x}) \nabla T(\bm{x}) \cdot \bm{e}_x~d\bm{x},
\end{equation}
where $\bm{e}_x$ is the unit vector in the horizontal direction and $|\Omega|$ is the area of the domain. Analogously, $\kappa_v$ is computed by imposing a vertical gradient ($T = 0$ on the bottom boundary, $T = 1$ on the top) and using the vertical unit vector $\bm{e}_y$ in Eq.~\eqref{eq:kappa_h}. We notice from \eqref{eq:kappa_h} that $\kappa_{eff}$ depends on both the microstructure $\mu$ and the solution $T$. This makes the inference with respect to $\mu$ has also to be carried out in problems \textbf{(P1)} and \textbf{(P3)} (see Eq.~\eqref{eq:p1_joint} and Eq.~\eqref{eq:p3_joint}). To mitigate this, the decoder $p_{\bt_u}(u|\bm{\beta})$ is designed to predict the flux field $\mu(\bm{x}) \nabla T(\bm{x})$ directly, rather than the temperature field $T(\bm{x})$. Hence, in problems \textbf{(P1)} and \textbf{(P3)}, the response field $u$ represents the flux.
By contrast, in problem \textbf{(P2)}, the design target is the temperature distribution itself, so the response field $u$ is the temperature field $T(\bm{x})$.

\subsubsection{Data generation and model setups}
\label{sec:gen_data}
The model architectures for both methods are described in Section~\ref{sec:network}. For training, we generate a dataset of $N = 10,000$ samples, each consisting of a microstructure $\hat{\mu}$, its associated flux or temperature field $\hat{\bm{u}}$ (on regular $64\times 64$ grids $\Xi_u$), and virtual observables $\hat{\bm{R}}$ containing $M = 100$ weighted residuals. Effective thermal conductivities $\kappa_{\mathrm{eff}}$ are also computed via the SP link described in Section \ref{sec:experiment_sp} for training PoreFlow. An additional $N_{\mathrm{test}} = 1000$ samples from the same distribution are used for evaluation.
Unless otherwise stated, we set the latent dimension to $d_{\beta} = 128$ for both methods. Training is performed using the ADAM optimizer with an initial learning rate of $5\times 10^{-4}$, halved every $200$ epochs. We use a batch size of $25$ and train for $1000$ epochs to ensure convergence. For the proposed model, we set $\lambda_{\mathrm{pde}} = 0.25$, $\lambda_{data} = 0.5$, and $\lambda_{kl} = 2$. For PoreFlow, the regression loss on effective properties is weighted by $1$, while the microstructure reconstruction loss is weighted by $2$. All experiments are conducted under identical hardware conditions on a 64-core AMD Ryzen CPU with an NVIDIA RTX 4090 GPU.

\subsection{Inverse design with target properties}
We first assess the proposed method on the inverse design problem \textbf{(P1)}, where the objective is to generate microstructures whose effective thermal conductivities lie within a target property region $\mathcal{T}_d$. Two scenarios are considered:
\begin{itemize}
    \item \textbf{Case 1:} the target region \textbf{overlaps} with the training distribution.
	\item \textbf{Case 2:} the target region lies \textbf{outside} the training distribution.
\end{itemize}
This comparison enables a joint assessment of the model’s interpolation accuracy within the training domain and its robustness in generalizing to property ranges beyond those observed during training.

\subsubsection{Forward prediction performance}
\begin{figure}[ht]
    \centering  
    \subfigure[True vs. Predicted $\kappa_{eff}=(\kappa_h,\kappa_v)$ with Design-GenNO]{\label{fig:fwd_our}
        \includegraphics[width=0.78\textwidth]{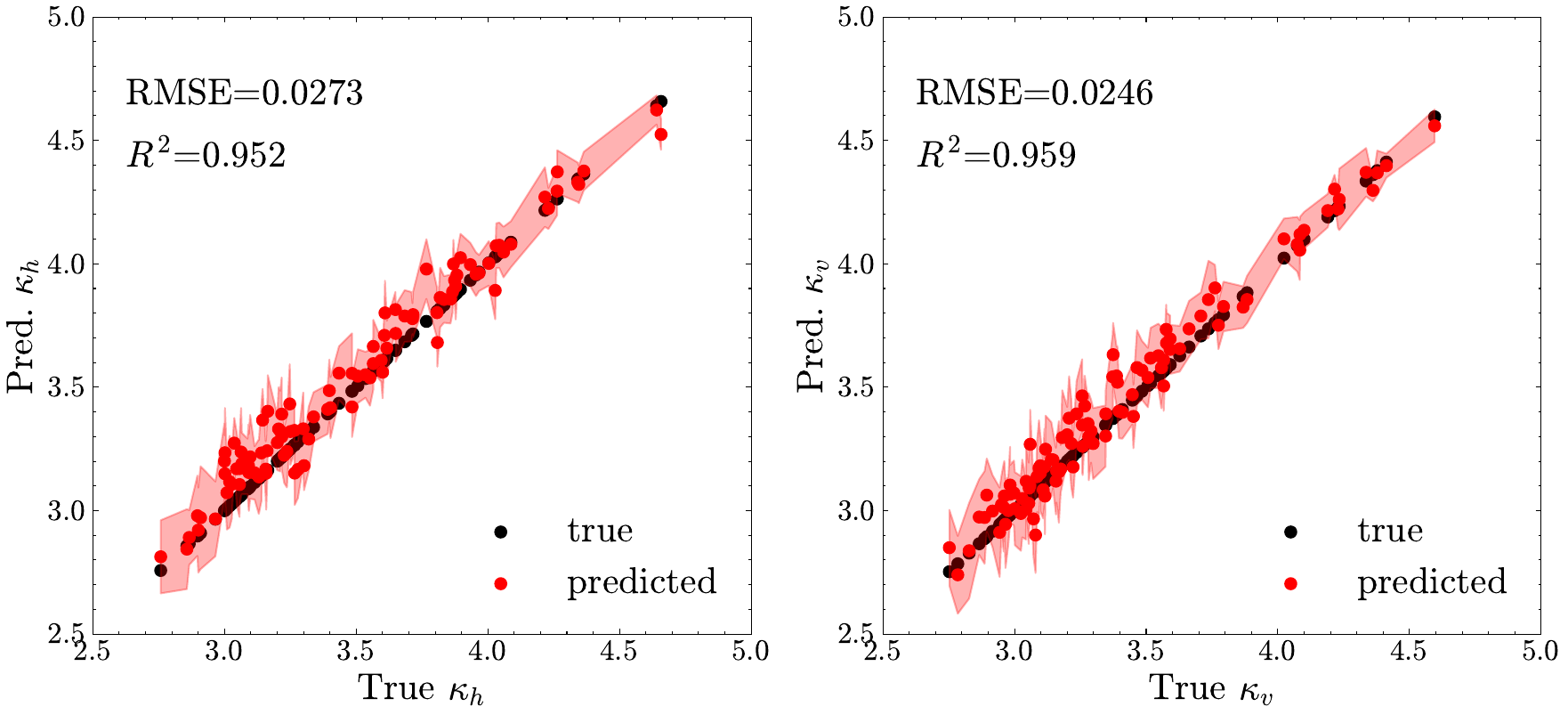}}
    \subfigure[True vs. Predicted $\kappa_{eff}=(\kappa_h,\kappa_v)$ with PoreFlow]{\label{fig:fwd_cmp}
        \includegraphics[width=0.78\textwidth]{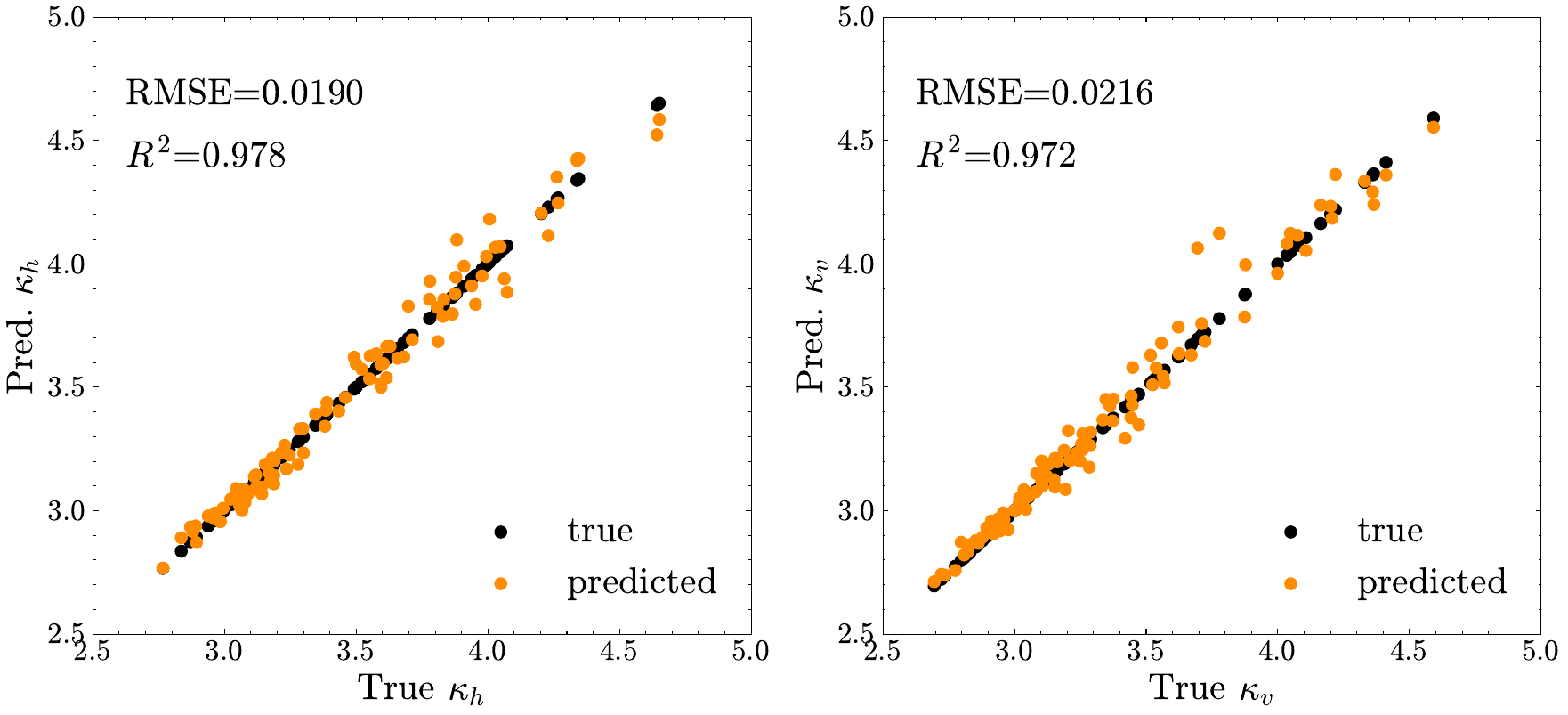}}
    \subfigure[Microstructure samples in testing dataset]{\label{fig:fwd_img}
        \includegraphics[width=0.99\textwidth]{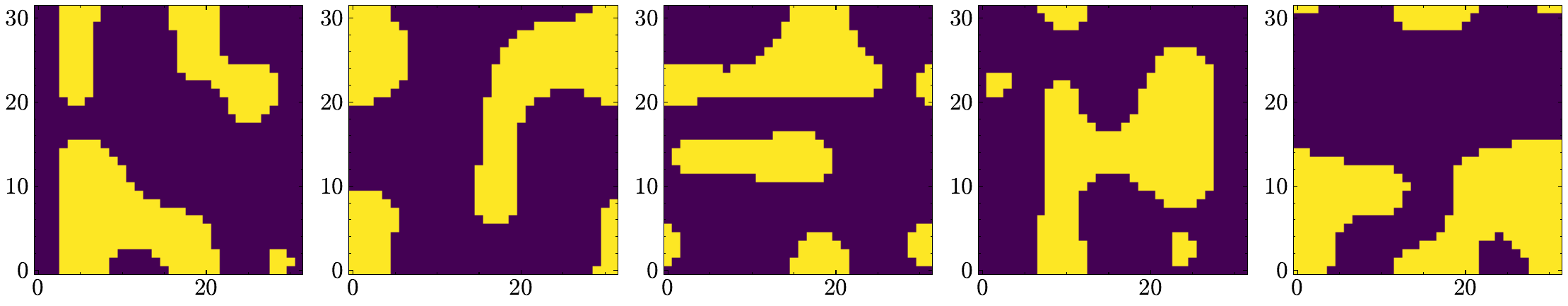}}
    \vspace{-0.25cm}
    \caption{Predictions of the effective thermal conductivity $\kappa_{eff} = (\kappa_h, \kappa_v)$, where $\kappa_h$ denotes the horizontal component and $\kappa_v$ the vertical component (for clarity, $100$ randomly selected test samples are shown): (a) predictions by the proposed Design-GenNO; (b) predictions by the PoreFlow; (c) five samples of microstructure in testing dataset (phase 1 represented by yellow color).}
    \label{fig:fwd}
\end{figure}

Before addressing the inverse design tasks, we first evaluate the predictive performance of the trained model on forward problems, aiming to verify that it accurately captures the underlying physics of the structure–property (SP) relationship.  In Figure~\ref{fig:fwd}, we present the predictions of the effective thermal conductivity $\kappa_{eff} = (\kappa_h, \kappa_v)$ on the testing dataset. For reference, Figure~\ref{fig:fwd_img} illustrates five microstructures sampled from the testing dataset (with phase 1 represented by yellow color).
The true versus predicted effective thermal conductivity by the proposed method on the testing set is plotted in Figure~\ref{fig:fwd_our}. The predicted mean effectivities are present with red dots for the proposed Design-GenNO, while the truths are present with black dots. For both $\kappa_h$ and $\kappa_v$, the points are tightly clustered around the $y = x$ line, indicating strong agreement between predicted means and ground truth. Quantitatively, the Root Mean Squared Errors (RMSE) are small, and the $R^2$ scores are close to unity for both $\kappa_h$ and $\kappa_v$, as reported in the same figure. Furthermore, the probabilistic nature of our model enables us to provide a $95\%$ trust region for each conductivity component. Most ground-truth values fall within these intervals, highlighting the reliability of the uncertainty estimates.
This uncertainty quantification is particularly valuable for inverse design with out-of-distribution targets \cite{wang_uncertainty-aware_2025}. In such cases, the model must explore regions that are far from the training space. A probabilistic model that faithfully represents predictive uncertainty can (i) signal when it is operating in less certain regimes, and (ii) guide sampling and optimization toward regions where the target is achievable while maintaining confidence in the prediction. As we will demonstrate in the following sections, this feature plays a crucial role in achieving successful designs beyond the training distribution.

For comparison, Figure~\ref{fig:fwd_cmp} shows the corresponding predictions of effective conductivity by  PoreFlow \cite{mirzaee2025inverse}. While it achieves even better RMSE and $R^2$ values and its predictions also align with the $y = x$ line, it differs fundamentally in its approach: PoreFlow directly maps the latent representation of the microstructure to the corresponding effective property without modeling the underlying physics. Moreover, it lacks probabilistic outputs, limiting its ability to identify and adapt to uncertainty in uncharted regions of the property space. This shortcoming will be evident in out-of-distribution design tasks.

\subsubsection{Case 1: Target region overlapping with training distribution}
\label{sec:case1}
\begin{figure}[ht]
    \centering  
    \subfigure[Distribution of $k_{eff}$ corresponding to designed microstructures by Design-GenNO]{\label{fig:design_case1_dist_our}
        \includegraphics[width=0.4\textwidth]{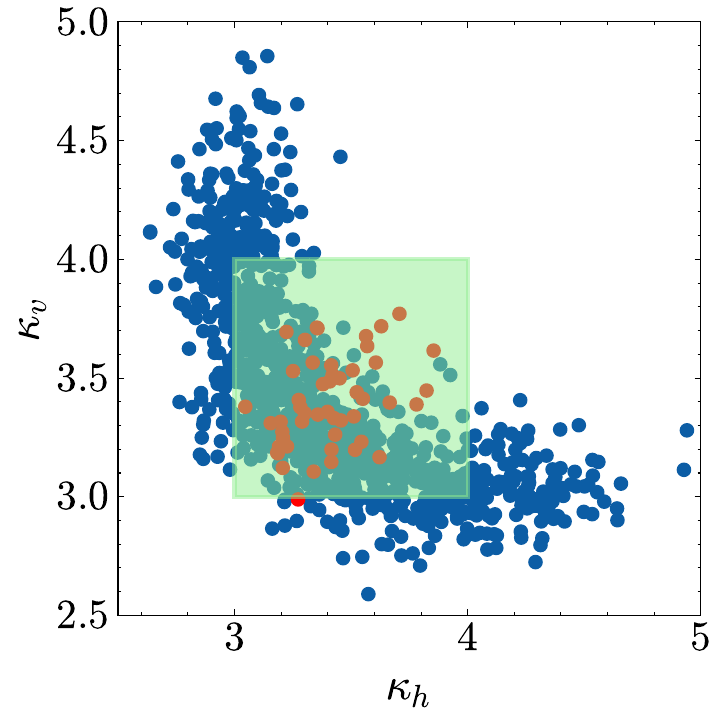}}
    \subfigure[Distribution of $k_{eff}$ corresponding to designed microstructures by PoreFlow]{\label{fig:design_case1_dist_cmp}
        \includegraphics[width=0.4\textwidth]{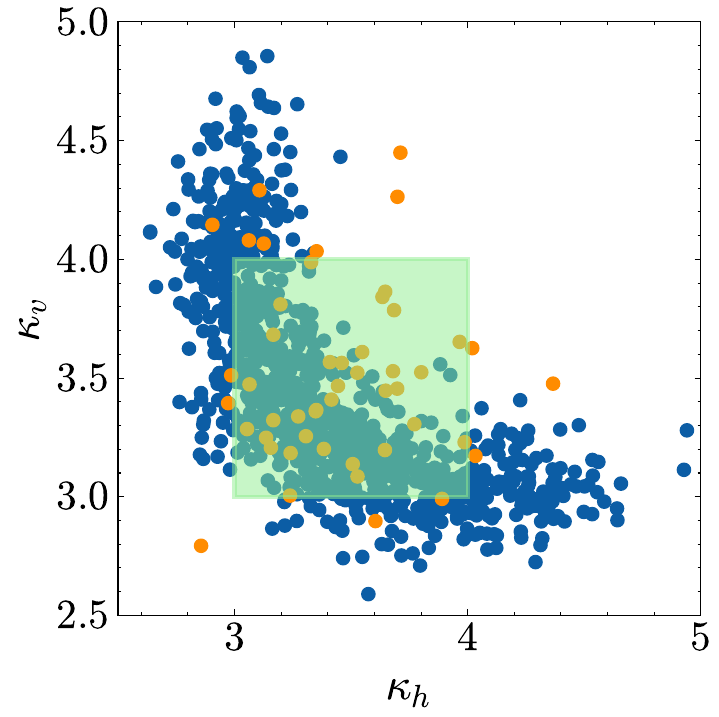}}
    \subfigure[Designed microstructure samples with the proposed Design-GenNO]{\label{fig:design_case1_img_our}
        \includegraphics[width=1.\textwidth]{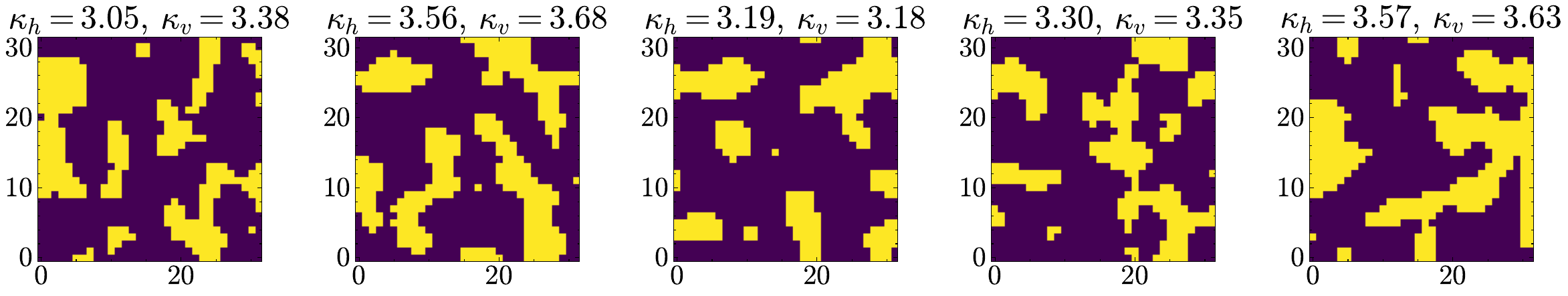}}
    \subfigure[Designed microstructure samples with the PoreFlow]{\label{fig:design_case1_img_cmp}
        \includegraphics[width=1.\textwidth]{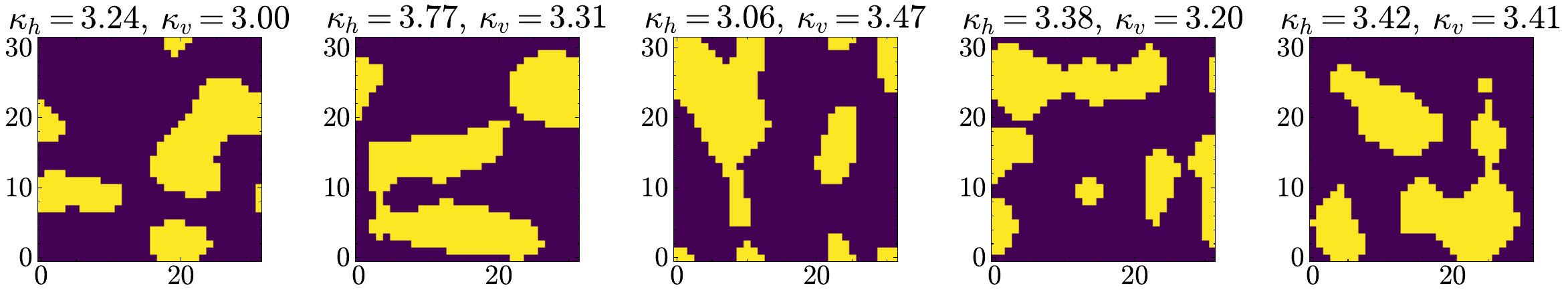}}
    \vspace{-0.25cm}
    \caption{Inverse microstructure design targeting the region $\mathcal{T}_d = [3, 4] \times [3, 4]$ (rectangle region in light green): (a) Effective thermal conductivity distribution of the testing dataset (blue dots) and the designed samples from Design-GenNO (red dots and $98\%$ are located in $\mathcal{T}_d$); (b) Effective thermal conductivity distribution of the testing dataset (blue dots) and the designed samples from PoreFlow (orange dots and $68\%$ are located in $\mathcal{T}_d$); (c) Five examples of designed microstructures from Design-GenNO; (d) Five examples of designed microstructures from PoreFlow.}
    \label{fig:design_case1}
\end{figure}
In the first scenario, we target the property region $\mathcal{T}_d = [3, 4] \times [3, 4]$ in the $(\kappa_h, \kappa_v)$ property space (Figure~\ref{fig:design_case1}). This region lies in the lower-left part of the property space and overlaps with the training dataset. The goal here is to design microstructures with moderate effective conductivities in both horizontal and vertical directions.

Following the strategy in Section~\ref{sec:sol_p1}, we define the distance as the squared Euclidean norm $
d(\bm{\beta}) = \|\kappa_{eff}(\mathcal{G}_{\bt^*_u}(\bm{\beta})) - \bar{\kappa}_{eff}\|^2$, where $\bar{\kappa}_{eff}$ is the center of $\mathcal{T}_d$, and set the tempering parameter $\tau=10$. Using Design-GenNO, we generated $50$ microstructures and evaluated their ground-truth properties. As shown in Figure~\ref{fig:design_case1_dist_our}, $98\%$ of the generated samples fall within the target region (red dots). Five representative examples are provided in Figure~\ref{fig:design_case1_img_our}. These results demonstrate that the proposed method not only achieves high accuracy in meeting target specifications but also produces diverse microstructures with consistent target properties, highlighting its strong interpolation ability.

For comparison, we applied the PoreFlow method with the $\bar{\kappa}_{eff}$ as conditions in the CNF model and generated another $50$ microstructures. The property distribution of these designs is shown in Figure~\ref{fig:design_case1_dist_cmp}, where $68\%$ of the samples fall inside the target region (orange dots). While the PoreFlow method can generate a fraction of valid microstructures, its success rate is significantly lower than that of Design-GenNO, indicating weaker performance and less reliable satisfaction of design constraints. Representative examples of PoreFlow-generated microstructures are shown in Figure~\ref{fig:design_case1_img_cmp}.
\subsubsection{Case 2: Target region outside training distribution}
\label{sec:case2}
\begin{figure}[ht]
    \centering  
    \subfigure[Distribution of $k_{eff}$ corresponding to designed microstructures by Design-GenNO]{\label{fig:design_case2_dist_our}
        \includegraphics[width=0.4\textwidth]{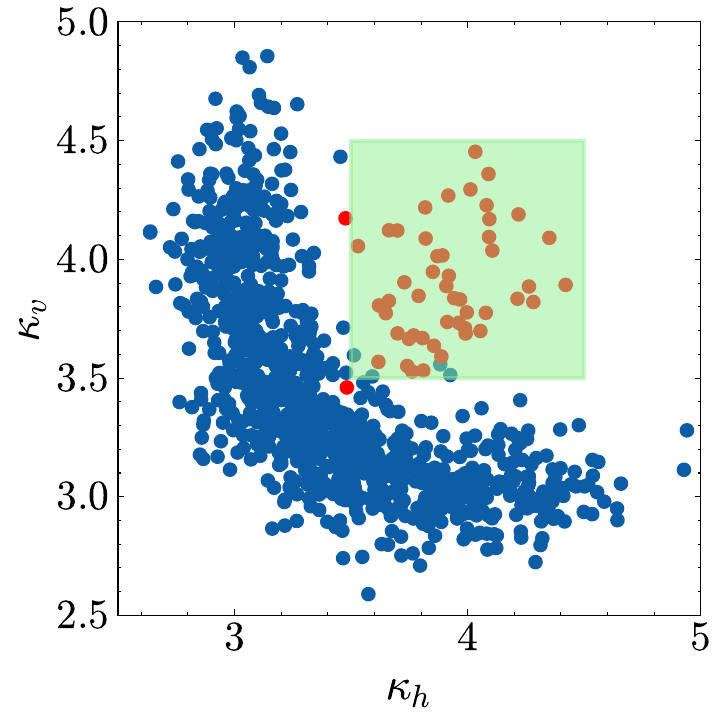}}
    \subfigure[Distribution of $k_{eff}$ corresponding to designed microstructures by PoreFlow]{\label{fig:design_case2_dist_cmp}
        \includegraphics[width=0.4\textwidth]{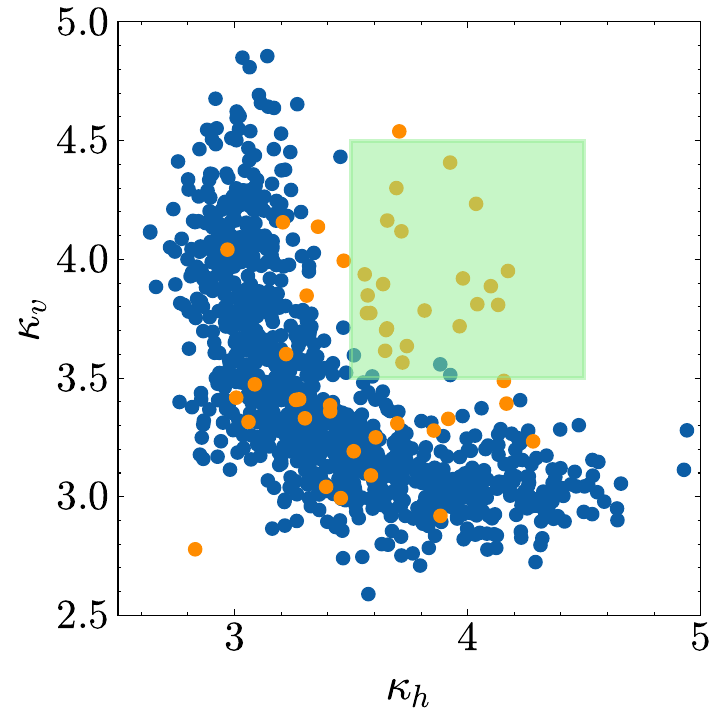}}
    \subfigure[Designed microstructure samples with the proposed Design-GenNO]{\label{fig:design_case2_img_our}
        \includegraphics[width=1.\textwidth]{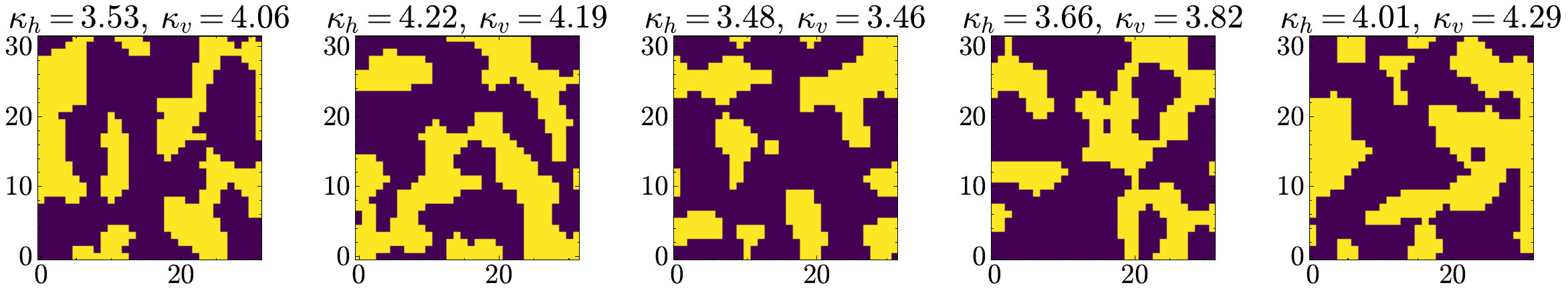}}
    \subfigure[Designed microstructure samples with the PoreFlow]{\label{fig:design_case2_img_cmp}
        \includegraphics[width=1.\textwidth]{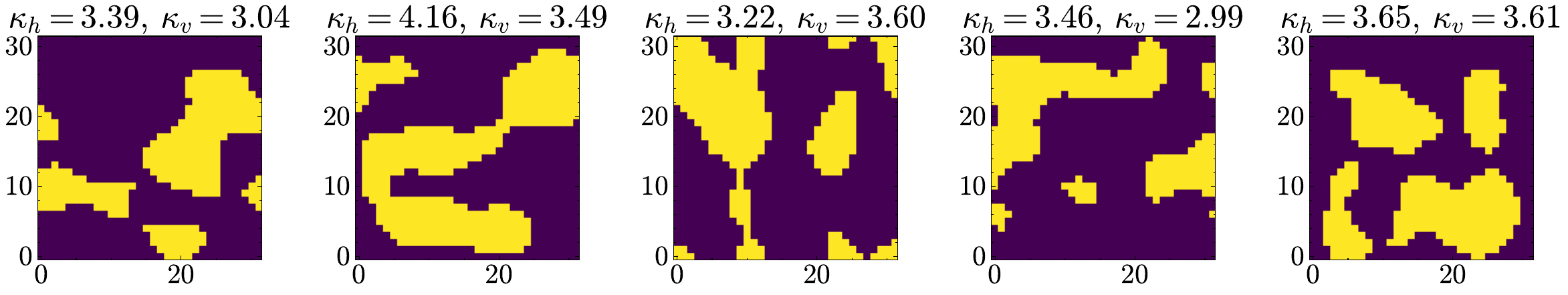}}
    \vspace{-0.25cm}
    \caption{Inverse microstructure design targeting the region $\mathcal{T}_d = [3.5, 4.5] \times [3.5, 4.5]$ (rectangle region in light green): (a) Effective thermal conductivity distribution of the testing dataset (blue dots) and the designed samples from Design-GenNO (red dots and $96\%$ are located in $\mathcal{T}_d$); (b) Effective thermal conductivity distribution of the testing dataset (blue dots) and the designed samples from PoreFlow (orange dots and $44\%$ are located in $\mathcal{T}_d$); (c) Five examples of designed microstructures from Design-GenNO; (d) Five examples of designed microstructures from PoreFlow.}
    \label{fig:design_case2}
\end{figure}
In the second scenario, the target property region is
$\mathcal{T}_d = [3.5, 4.5]\times [3.5, 4.5]$, corresponding to the light green rectangular area in Figure~\ref{fig:design_case2}. Since this region lies entirely outside the property range covered by the training dataset, this task serves as a stringent evaluation of the model’s ability to extrapolate and generate microstructures with effective conductivities not encountered during training. The results in this case reveal the benefits of combining physics-informed modeling with probabilistic predictions for robust out-of-distribution performance.

Following the same strategy  as in Case 1, we generated $50$ microstructures using the proposed Design-GenNO and computed their corresponding ground-truth properties. The resulting property distribution (red dots  in Figure~\ref{fig:design_case2_dist_our}) shows that $96\%$ of the generated samples fall within the target region. Five representative designs are shown in Figure \ref{fig:design_case2_img_our}. Compared to the microstructure samples used for training (blue dots), the generated microstructures exhibit notable morphological differences: boundaries between the two phases are less smooth, and the phase domains are more spatially dispersed. These morphological changes increase thermal conductivity in both horizontal and vertical directions while preserving other features. This demonstrates the model’s ability to modify microstructural features in a targeted manner to achieve desired properties, even for regions absent from the training region.

For comparison, we applied the PoreFlow method to generate another $50$ microstructures with the same property targets. The ground-truth effective thermal conductivity corresponding to the generated microstructures (orange dots in Figure \ref{fig:design_case2_dist_cmp}) reveals that only $44\%$ of the samples meet the target specification. Five representative PoreFlow-generated designs are shown in Figure \ref{fig:design_case2_img_cmp}, which visually resemble the microstructures present in the training space (see Figure \ref{fig:fwd_img}). This suggests that PoreFlow tends to replicate known morphologies, limiting its ability to explore new structural patterns necessary for achieving out-of-distribution properties. 

\subsection{Inverse design with target temperature field}
\begin{figure}[t]
    \centering  
    \subfigure[Target field $T_{target}$]{\label{fig:design_field_obs}
        \includegraphics[width=0.35\textwidth]{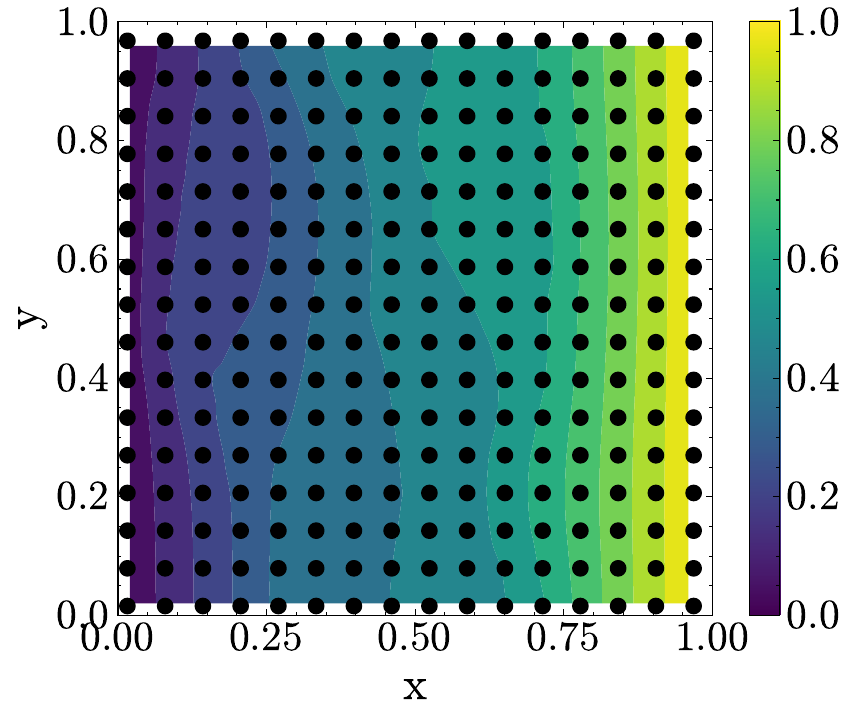}}
    \subfigure[True $\mu_{target}$]{\label{fig:design_field_true}
        \includegraphics[width=0.3\textwidth]{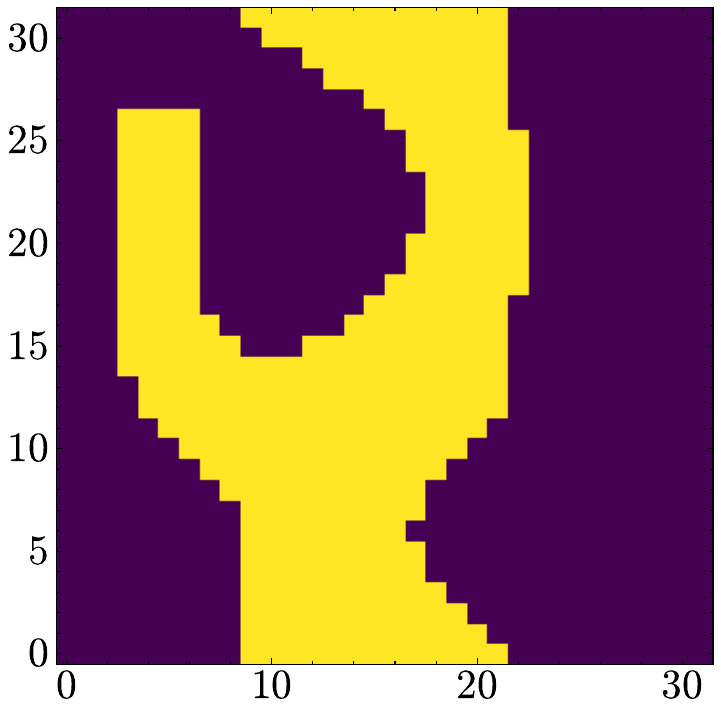}}
    \subfigure[Recovered $\mu_{rec}$]{\label{fig:design_field_pred}
        \includegraphics[width=0.3\textwidth]{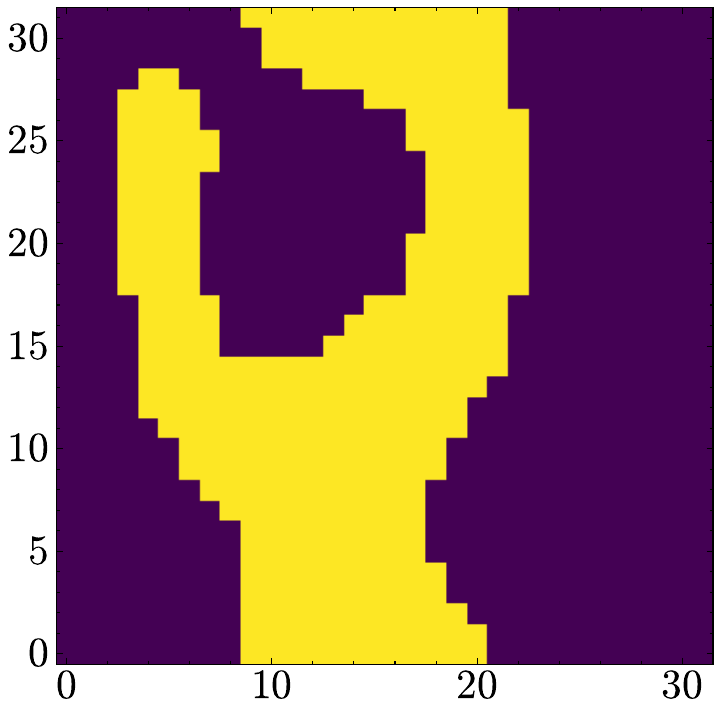}}
    \vspace{-0.25cm}
    \caption{Inverse microstructure design targeting a temperature field with the proposed Design-GenNO: (a) The desired temperature target $T_{target}$ on a set of sensors $\Xi_T$ (dots in black color); (b) The true microstructure $\mu_{target}$ used to generate $T_{target}$ (c) The recovered microstructure $\mu_{rec}$ with Design-GenNO (with $I_{corr}=0.951$).}
    \label{fig:design_field}
\end{figure}
In this example, we evaluate the proposed method on design problem \textbf{(P2)}, which targets a specific temperature profile. Many competing models are trained on a fixed set of properties, so when design objectives fall outside this range, re-training is typically required. In contrast, Design-GenNO learns the entire PDE solution field, enabling it to accommodate arbitrary property targets without the need for re-training.

To construct the target, we first sample a microstructure $\mu_{target}$ (see Figure \ref{fig:design_field_true}) and compute the corresponding temperature field $T_{target}$ by solving the PDE~\eqref{eq:pde} with Dirichlet boundary conditions $T = 1$ at $x = 1$ and $T = 0$ elsewhere. The target is defined as the measurements of $T_{target}$ at a set of sensors $\Xi_T = (\bm{\xi}_1, \dots, \bm{\xi}_m)$, i.e., $T_{target}(\Xi_T)$ (see Figure \ref{fig:design_field_obs}). This task can thus be viewed as recovering $\mu_{target}$ from finite temperature measurements $T_{target}(\Xi_T)$. Since the PoreFlow method cannot accommodate field response targets, we evaluate only the proposed method in this case.

Following the inverse design strategy described in Section~\ref{sec:sol_p2} for problem \textbf{(P2)} (with $\tau_u^{-1}$ is selected as $\tau_u^{-1}=\|T_{target}(\Xi_T)\|^2$), we recover the microstructure using the proposed Design-GenNO, which is denoted as $\mu_{rec}$. The result is shown in Figure \ref{fig:design_field_pred}. Visually, the recovered design closely matches the target, indicating that the proposed method can effectively reconstruct microstructures from indirect field measurements.
To quantify the similarity, we compute the cross-correlation indicator \cite{bourke1996cross}:
\be 
I_{corr} = \frac{\sum_i \tilde{\mu}_{target}^2(\bm{\xi}_i)\tilde{\mu}_{rec}^2(\bm{\xi}_i)}{\sqrt{\sum_i \tilde{\mu}_{target}^2(\bm{\xi}_i)}\sqrt{\sum_i \tilde{\mu}_{rec}^2(\bm{\xi}_i)}}.
\ee  
which ranges from $0$ (no correlation) to $1$ (perfect match), where $\tilde{\mu}$ denotes the scaled microstructure, obtained by normalizing the values of $\mu$ to the interval $[0,1]$. The recovered design achieves $I_{corr} = 0.951$, confirming strong agreement between the reconstructed and target microstructures.

\subsection{Maximization design problem without labeled training pairs}
\label{sec:maxdesign}
In this example, we test the proposed method on a more challenging design objective: finding microstructures that maximize a physical property. Such maximization problems are common in topology and material design \cite{feng2024topology, huang2015inverse}.  In this example, we target the expectation ratio between horizontal and vertical effective conductivities, namely $\frac{\kappa_h}{\kappa_v}$. We denote with $F_{\kappa_{eff}}(u)$ the known function that yields the ratio of interest with respect to the response field $u$. 

We note first that in this test, the proposed generative model was trained solely using pairs of microstructures $\hat{\mu}$ and virtual observables $\hat{\mathbf{R}}$, i.e., without any labeled data-pairs $\mathcal{D}_{l}$. This was achieved by simply removing the likelihood corresponding to real observables \eqref{eq:likel} and discarding the $\mathcal{L}_{l}$ term from the training loss \eqref{eq:loss}. The purpose of this setup is to demonstrate that the proposed framework can be trained in a purely physics-driven manner, guided solely by the governing PDE and virtual observables, while still enabling high-performance inverse microstructure design.

The proposed method is well-suited to this type of target for two reasons. First, by directly capturing the latent physics of the SP linkage through DNOs, it can handle general design objectives derived from PDE solutions—even when those PDE solutions were never explicitly used in training. Second, the differentiable MultiONet architecture allows for efficient gradient-based optimization by computing sensitivities of the output with respect to the latent variables $\bmb$  in a single pass.  As in the previous cases, therefore, the optimization/inference can be carried out in the finite and lower-dimensional $\bmb$-space and the results can be readily propagated through the microstructural decoder in order to obtain the corresponding microstructures.

\begin{figure}[ht]
    \centering  
    \subfigure[Distribution of $k_{eff}=(\kappa_h,\kappa_v)$]{\label{fig:design_maximize_dist}
        \includegraphics[width=0.4\textwidth]{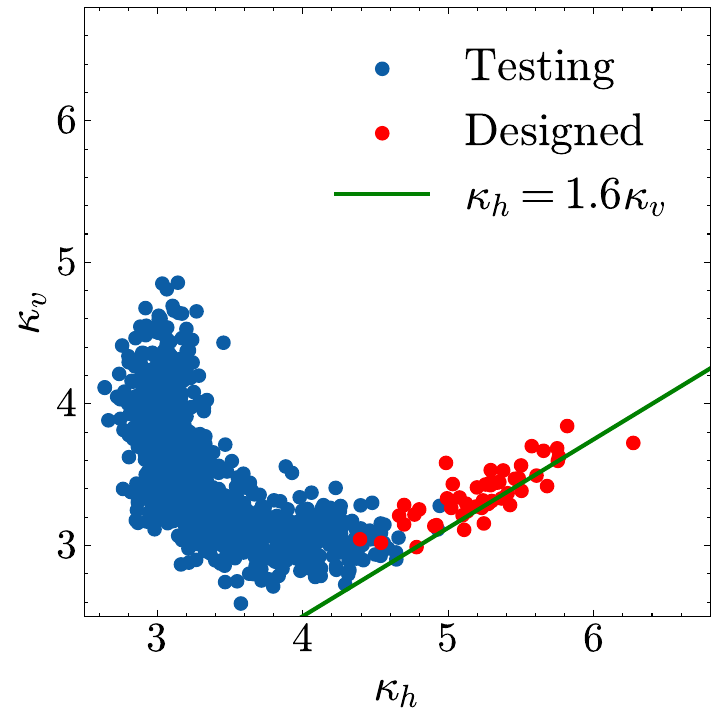}}
    \subfigure[Distribution of $\kappa_h/\kappa_v$ ratio]{\label{fig:design_maximize_ratio}
        \includegraphics[width=0.415\textwidth]{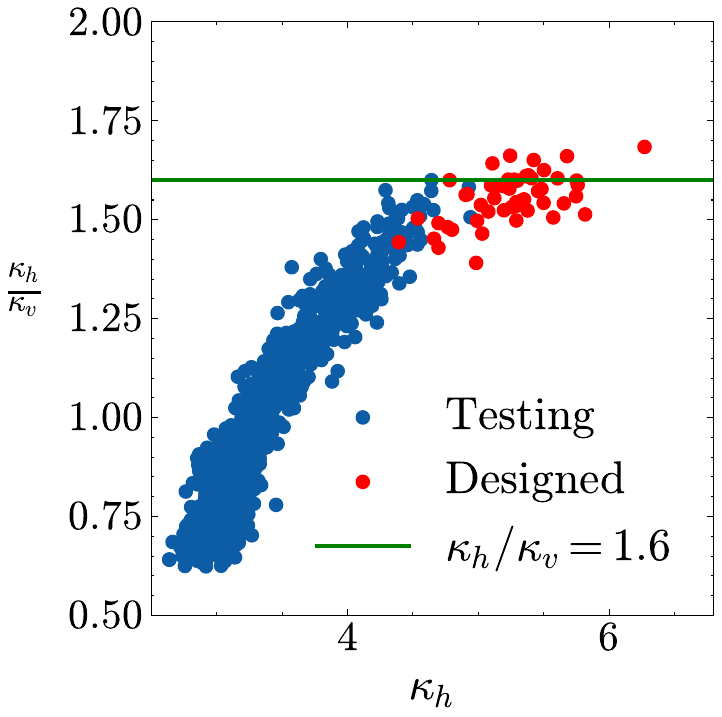}}
    \subfigure[Designed microstructure samples with the proposed Design-GenNO]{\label{fig:design_maximize_img}
        \includegraphics[width=1.\textwidth]{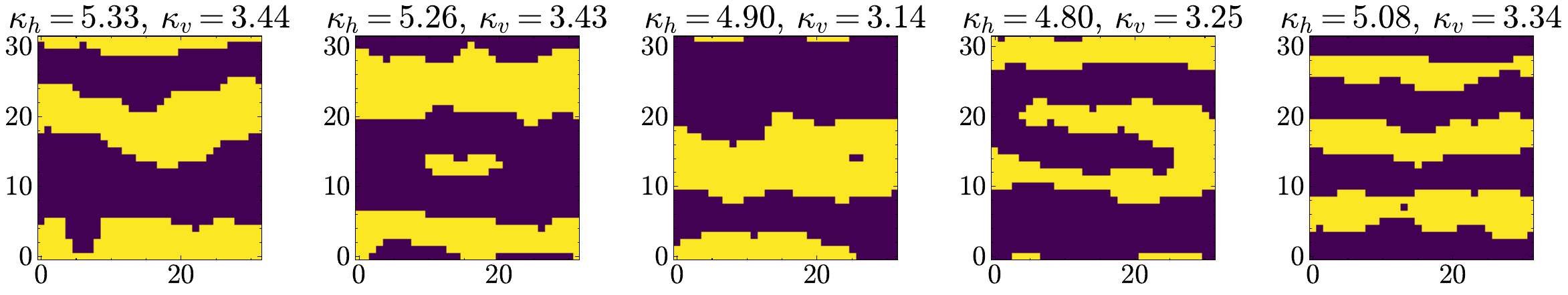}}
    \vspace{-0.25cm}
    \caption{
    Inverse microstructure design targeting maximization of $F(\kappa_{eff})$ with the proposed Design-GenNO: (a) Effective thermal conductivity distribution of the testing dataset (blue dots) and the designed samples from Design-GenNO (red dots); (b) The distribution of $\kappa_h/\kappa_v$ ratio of the testing dataset (blue dots) and the designed samples from Design-GenNO (red dots); (c) Five examples of designed microstructures from Design-GenNO.
    }
    \label{fig:design_maximize}
  
\end{figure}
Using this strategy in Section \ref{sec:sol_p3}, we generated $50$ posterior samples $\bm{\beta}$ and decoded them into microstructures. Their corresponding effective properties $\kappa_{\text{eff}} = (\kappa_h, \kappa_v)$ are shown in Figure \ref{fig:design_maximize_dist}. The points cluster in the lower-right region of the property space (high $\kappa_h$, low $\kappa_v$), indicating that the designs achieve the desired anisotropic behavior. Notably, most designs lie beyond the range of properties in the training distribution, further demonstrating the method’s capacity to produce out-of-distribution microstructures—an essential feature for discovering novel material configurations. 
The distribution of conductivity ratios $\kappa_h/\kappa_v$ is shown in Figure~\ref{fig:design_maximize_ratio}, where most values lie in the upper-right region, corresponding to both high $\kappa_h$ and large ratios. Several microstructures achieve ratios well beyond those observed in training distribution (e.g., $\frac{\kappa_h}{\kappa_v}>1.6$), further confirming out-of-distribution generalization of the proposed method.
Five designed microstructures are shown in Figure \ref{fig:design_maximize_img}. These designs exhibit phase arrangements that are more continuous and connected along the horizontal direction, while more fragmented vertically. This structural anisotropy gives rise to a high $\kappa_h / \kappa_v$ ratio, fulfilling the intended design objective.

\section{Concluding remarks}
\label{sec:conclusion}
In this work, we proposed Design-GenNO, a generative neural operator framework for inverse microstructure design with application in two-phase materials. The goal is to design microstructures that satisfy user-specified design targets, such as macroscopic effective properties or microscopic field responses. Unlike existing PIML or DNO-based approaches, which either struggle with microstructure-centered design or rely on numerous expensive labeled training data, Design-GenNO combines the strengths of generative modeling, operator learning, and physics-informed training to address these challenges in a principled and efficient manner.

The proposed framework leverages several key innovations. First, it employs a probabilistic generative latent space to compactly represent microstructures while naturally enabling uncertainty quantification. Second, MultiONet-based decoders map latent variables to both microstructures and response fields, supporting functional outputs beyond finite-dimensional mappings. Third, physics-informed training with virtual observables from governing PDEs reduces reliance on costly labeled data. Finally, a normalizing flow prior provides high-quality initializations for gradient-based optimization in the latent space, improving sampling efficiency, avoiding poor local minima, and accelerating convergence. Collectively, these innovations allow Design-GenNO to efficiently explore design spaces, optimize in a lower-dimensional and well-structured latent space, and deliver robust predictions even in out-of-distribution regimes.

We validated the proposed framework on a series of inverse design tasks with increasing difficulty, including property-matching problems, recovery of microstructures from temperature measurements, and maximization of conductivity ratios. Across all tasks, Design-GenNO consistently achieved high accuracy, generated diverse and physically meaningful microstructures, and significantly outperformed the state-of-the-art baseline PoreFlow. Notably, the framework demonstrated strong extrapolation capability by successfully designing microstructures that achieve effective properties outside the training distribution, underscoring its robustness and generality.

Looking forward, several promising directions remain. A natural extension is to incorporate the full processing–structure–property (PSP) linkage into the generative model \cite{generale2023bayesian,zang2025psp}, enabling goal-oriented design that integrates processing conditions rather than focusing solely on microstructures. Another direction is to extend the framework to more complex SP linkage objectives, such as nonlinear property–response curves \cite{bastek_inverse_2023}. Addressing these challenges would further broaden the applicability of the framework and establish Design-GenNO as a versatile tool for materials discovery and inverse design.

\section*{Acknowledgement}
Funded by the Excellence Strategy of the Federal Government and the L\"ander in the context of the ARTEMIS Innovation Network.

\newpage
\bibliographystyle{elsarticle-num}
\bibliography{ref.bib}

\appendix
\section{The MultiONet architecture}
\label{sec:model_for_multionet}
The MultiONet employs a separable representation consisting of two subnetworks:
\begin{itemize}
	\item \textbf{Trunk network}: Encodes the spatial coordinates $\mathbf{x} \in \Omega$ of the output field.
	\item \textbf{Branch network}: Extracts features from the input latent vector $\bm{\beta}$.
\end{itemize}
It differs from the original DeepONet in that it aggregates information from multiple intermediate layers of both subnetworks. Specifically, the final output is obtained by averaging weighted inner products between the layerwise branch and trunk features. This design enhances predictive accuracy without increasing the number of trainable parameters.
Formally, the MultiONet mapping $\mathcal{G}(\bm{\beta})(\mathbf{x})$ is expressed as:
\begin{equation}\label{eq:multionet}
\mathcal{G}(\bm{\beta})(\mathbf{x}) =
\frac{1}{l} \sum_{k=1}^{l} w^{(k)}
\left( b^{(k)}(\bm{\beta}) \odot t^{(k)}(\mathbf{x}) \right) + b_0,
\end{equation}
where $b^{(k)}(\bm{\beta})$ and $t^{(k)}(\mathbf{x})$ denote the outputs from the $k$-th layers of the branch and trunk networks, $l$ represents the total number of layers, $w^{(k)}$ indicates trainable weight, $b_0$ is the bias term, and $\odot$ represents the inner product operation.

\section{Details of the loss}
\label{sec:loss_terms}
\paragraph{\textbf{The loss term $\mathcal{L}_u(\bt,\bm{\eta})$}}
For unlabeled data $\mathcal{D}_{u}$, the ELBO term in \eqref{eq:likeuelbo} is:
\be
\begin{array}{ll}
\mathcal{L}_u(\bt,\bm{\eta}) &= \sum_{n_u=1}^{N_u}   \left<  \log \cfrac{p_{\bt_\mu}(\hat{\mu}^{(n_u)} |\bmb^{(n_u)})~p_{\bt_{\bmb}} (\bmb^{(n_u)})}{q_{\bm{\eta}}(\bmb^{(n_u)} | \hat{\mu}^{(n_u)})   } \right>_{q_{\bm{\eta}}(\bmb^{(n_u)} | \hat{\mu}^{(n_u)})} \\
& = \sum_{n_u=1}^{N_u} \mathcal{L}^{(n_u)}_u(\bt, \bm{\eta}) 
\end{array}
\ee
where $\mathcal{L}^{(n_u)}_u(\bt,\bm{\eta})$ denotes the per-sample contribution. Dropping the data index $(n_u)$ for clarity, we decompose it as:
\be
\begin{array}{ll}
\mathcal{L}^{(n_u)}_u(\bt,\bm{\eta}) & =  \left<\log p_{\bt_\mu}(\hat{\mu} |\bm{\beta})\right>_{q_{\bm{\eta}}(\bs{\beta}|\hat{\mu})} + \left<\log \frac{p_{\bt_{\beta}}(\bs{\beta})}{q_{\bm{\eta}}(\bs{\beta}|\hat{\mu})}\right>_{q_{\bm{\eta}}(\bs{\beta}|\hat{\mu})} \\
& = \mathcal{L}_{rec}(\bt,\bm{\eta}) + \mathcal{L}_{kl}(\bt, \bm{\eta}).
\end{array}
\ee
Hence, it consists of two terms: $\mathcal{L}_{rec}$ representing microstructure reconstruction loss, and $\mathcal{L}_{kl}$ representing KL regularization that enforces prior matching.
For two-phase microstructures, the Multivariate Bernoulli model is adopted for $p_{\bt_\mu}(\hat{\mu}|\bmb)$ as in \refeqp{eq:mu_pred}, leading to:
\begin{equation}\label{eq:loss_rec}
\mathcal{L}_{rec}(\bm{\theta},\bm{\eta})
= \sum_{j=1}^{J} \left[ z_j \log \sigma\big(\mathcal{G}_{\bm{\theta}_\mu}(e_{\bm{\eta}}(\hat{\mu}))(\bm{x}_j) \big) + (1-z_j) \log\big(1 - \sigma(\mathcal{G}_{\bm{\theta}_\mu}(e_{\bm{\eta}}(\hat{\mu}))(\bm{x}_j) \big) \right],
\end{equation}
where $\sigma(\cdot)$ is the sigmoid function, and $z_j \in \{0,1\}$ indicates the material phase of $\hat{\mu}$ at $\bm{x}_j$.
The KL term encourages the approximate posterior to align with the prior distribution over $\bm{\beta}$. Since $q_{\bm{\eta}}$ is a delta distribution centered at $e_{\bm{\eta}}(\hat{\mu})$, the expectation simplifies to evaluation of the prior density:
$$
\mathcal{L}_{kl}(\bm{\theta},\bm{\eta})
= \log p_{\bm{\theta}_\beta}\big( e_{\bm{\eta}}(\hat{\mu}) \big) + \text{const}.
$$
With the prior $p_{\bm{\theta}_\beta}$ defined via the inverse RealNVP transformation \eqref{eq:NF}, its density is given by:
$$
p_{\bm{\theta}_\beta}(\bm{\beta})
= \mathcal{N}\!\left(f_{\bm{\theta}_\beta}(\bm{\beta});\, \mathbf{0},\mathbf{I}\right)
\cdot \left| \det J_{f_{\bm{\theta}_\beta}}(\bm{\beta}) \right|,
$$
where $J_{f_{\bm{\theta}_\beta}}$ is the Jacobian of $f_{\bt_\beta}$. Thus,
\begin{equation}\label{eq:loss_kl}
\mathcal{L}_{kl}(\bm{\theta},\bm{\eta})
= -\frac{1}{2} \left\| f_{\bm{\theta}_\beta}(e_{\bm{\eta}}(\hat{\mu})) \right\|_2^2
+ \log \left| \det J_{f_{\bm{\theta}_\beta}}(e_{\bm{\eta}}(\hat{\mu})) \right|
+\text{const}.
\end{equation}

In practice, we incorporate a weighting coefficient $\lambda_{kl}$ to modulate the KL-divergence term in the ELBO. This follows the well-established $\beta$-VAE strategy \cite{higgins2017beta}, which allows for balancing the trade-off between latent regularization and reconstruction fidelity. 

\paragraph{\textbf{The loss term $\mathcal{L}_l(\bt,\bm{\eta})$}} For the real observables $\mathcal{D}_l$, which contain both microstructures and their corresponding response fields, the ELBO contribution in \eqref{eq:likelelbo} is:
\be
\begin{array}{ll}
\mathcal{L}_l(\bt, \bm{\eta}) 
& =\sum_{n_l=1}^{N_l} \left( \left< \log  p(\hat{\bm{u}}^{(n_l)} | \mathcal{G}_{\bm{\theta}_u}(\bm{\beta}^{(n_l)}) ) p_{\bt_\mu}(\hat{\mu}^{(n_l)} | \bmb^{(n_l)})  \right>_{q_{\bm{\eta}}(\bmb^{(n_l)} | \hat{\mu}^{(n_l)})} \right. \\
&  \qquad  \qquad \left. \left< \log \cfrac{~p_{\bt_{\bmb}} (\bmb^{(n_l)})}{q_{\bm{\eta}}(\bmb^{(n_l)} | \hat{\mu}^{(n_l)})} \right>_{q_{\bm{\eta}}(\bmb^{(n_l)} | \hat{\mu}^{(n_l)})}  \right) \\
& = \sum^{N_l}_{n_l=1} \mathcal{L}^{(n_l)}_l(\bt, \bm{\eta}) 
\end{array}
\ee
where per-sample $\mathcal{L}^{(n_l)}_l(\bt, \bm{\eta})$ contribution can be decomposed as (index $(n_l)$ omitted for clarity):
\be
\begin{array}{ll}
\mathcal{L}^{(n_l)}_l(\bt,\bm{\eta}) & =  \left<\log p(\hat{\bm{u}} | \mathcal{G}_{\bt_u}(\bs{\beta})) \right>_{q_{\bm{\eta}}(\bs{\beta}|\hat{\mu})} + \left<\log p_{\bt_\mu
}(\hat{\mu} | \bm{\beta})\right>_{q_{\bm{\eta}}(\bs{\beta}|\hat{\mu})} \\
&\quad + \left<\log \frac{p_{\bt_{\beta}}(\bs{\beta})}{q_{\bm{\eta}}(\bs{\beta}|\hat{\mu})}\right>_{q_{\bm{\eta}}(\bs{\beta}|\hat{\mu})} \\
& = \mathcal{L}_{data}(\bt,\bm{\eta}) + \mathcal{L}_{rec}(\bt,\bm{\eta}) + \mathcal{L}_{kl}(\bt, \bm{\eta}).
\end{array}
\ee
Thus, this loss contains three contributions: $\mathcal{L}_{data}$ matches predicted response fields to observed fields; $\mathcal{L}_{rec}$ reconstructs the microstructure; and $\mathcal{L}_{kl}$ regularizes the latent distribution. With the Gaussian likelihood adopted in \refeqp{eq:ulike}, $\mathcal{L}_{data}$ can be expressed as a weighted $\ell_2$ form:
\begin{equation}
\mathcal{L}_{data}(\bm{\theta},\bm{\eta})
= - \frac{\lambda_{data}}{2}\, \big\| \hat{\bm{u}} - \mathcal{G}_{\bm{\theta}_u}(e_{\bm{\eta}}(\hat{\mu}))(\Xi_u) \big\|_2^2,
\end{equation}
up to an additive constant. The terms $\mathcal{L}_{rec}$ and $\mathcal{L}_{kl}$ retain the same definitions as in \eqref{eq:loss_rec} and \eqref{eq:loss_kl}.

\paragraph{\textbf{The loss term $\mathcal{L}_v(\bt,\bm{\eta})$}} For the virtual observables $\mathcal{D}_v$, which contain microstructures with physics-based virtual observables (derived from PDE residuals), the ELBO contribution in \eqref{eq:likevelbo} is:
\be
\begin{array}{ll}
\mathcal{L}_{v}(\bt,\bm{\eta})
& = \sum_{n_v=1}^{N_v} \left( \left< \log p\!\left(\hat{\bs{R}}^{(n_v)} \mid \mathcal{G}_{\bm{\theta}_u}(\bm{\beta}^{(n_v)}) , \hat{\mu}^{(n_v)}\right)\,
\right>_{q_{\bm{\eta}}(\bmb^{(n_v)} | \hat{\mu}^{(n_v)})} \right. \\
&\quad \left. + \left< \log  \cfrac{p_{\bt_\mu}\!\left(\hat{\mu}^{(n_v)} \mid \bmb^{(n_v)}\right)\,
p_{\bt_{\bmb}}\!\left(\bmb^{(n_v)}\right)\ }{q_{\bm{\eta}}(\bmb^{(n_v)} | \hat{\mu}^{(n_v)}) }  \right>_{q_{\bm{\eta}}(\bmb^{(n_v)} | \hat{\mu}^{(n_v)}) } \right) \\
& = \sum^{N_v}_{n_v}\mathcal{L}^{(n_v)}_v(\bt,\bm{\eta})
\end{array}
\ee
where per-sample $\mathcal{L}^{(n_v)}_v(\bt, \bm{\eta})$ contribution can be decomposed as (index $(n_v)$ omitted for clarity):
\be
\begin{array}{ll}
\mathcal{L}^{(n_v)}_v(\bt,\bm{\eta}) & =  \left<\log p(\hat{\bs{R}} |  \mathcal{G}_{\bt_u}(\bs{\beta}), \hat{\mu}) \right>_{q_{\bm{\eta}}(\bs{\beta}|\hat{\mu})} + \left<\log p_{\bt_\mu}(\hat{\mu} | \bm{\beta})\right>_{q_{\bm{\eta}}(\bs{\beta}|\hat{\mu})} \\
&\quad + \left<\log \frac{p_{\bt_{\beta}}(\bs{\beta})}{q_{\bm{\eta}}(\bs{\beta}|\hat{\mu})}\right>_{q_{\bm{\eta}}(\bs{\beta}|\hat{\mu})} \\
& = \mathcal{L}_{pde}(\bt,\bm{\eta}) + \mathcal{L}_{rec}(\bt,\bm{\eta}) + \mathcal{L}_{kl}(\bt, \bm{\eta}).
\end{array}
\ee
Thus, the loss includes three contributions: $\mathcal{L}_{pde}$ enforces physics constraints via PDE residual minimization; $\mathcal{L}_{rec}$ ensures faithful microstructure reconstruction; and $\mathcal{L}_{kl}$ regularizes the latent distribution against the prior. With the {\em virtual} likelihood defined in \refeqp{eq:vlike}, the PDE-informed loss term is expressed as:
\be
\label{eq:loss_pde}
\mathcal{L}_{pde}(\bm{\theta},\bm{\eta})
= - \frac{\lambda_{pde}}{2} \sum_{m=1}^M r_m^2\!\Big(\mathcal{G}_{\bm{\theta}_u}(e_{\bm{\eta}}(\hat{\mu})),\, \hat{\mu}\Big),
\ee
up to an additive constant. The $\mathcal{L}_{rec}$ and $\mathcal{L}_{kl}$ terms follow the same definitions as in \eqref{eq:loss_rec} and \eqref{eq:loss_kl}. 

\section{Model setups}
\label{sec:network}
In this section, we describe the neural-network architectures and hyperparameters used for the proposed Design-GenNO method and the baseline PoreFlow method in Section~\ref{sec:experiments}.
\subsection{The proposed Design-GenNO framework}
\label{sec:net_our}

\paragraph{The Encoder $e_{\bm{\eta}}$} The encoder $e_{\bm{\eta}}$ extracts latent representations from the input microstructure $\hat{\mu}$ using a feed-forward fully connected network (FFCN). The network has three hidden layers: the first layer flattens the input to a vector, followed by two dense layers with $512$ and $128$ neurons, respectively. The SiLU activation is applied to all hidden layers, while a Tanh activation is applied to the output layer.

\paragraph{The Decoder $\mathcal{G}_{\bt_\mu}$} The decoder $\mathcal{G}_{\bt_\mu}$ adopts the MultiONet architecture (Figure \ref{fig:MultiONet}). Both the branch and trunk networks are FFCNs with five hidden layers and $256$ neurons per layer. The trunk network employs a custom activation,
\begin{equation}\label{eq:silu_sin}
\text{SiLU\_Sin}(x) = \text{SiLU}(\sin(\pi x + \pi)) + x,
\end{equation}
while the branch network uses another custom activation,
\begin{equation}\label{eq:silu_id}
\text{SiLU\_Id}(x) = \text{SiLU}(x) + x.
\end{equation}
Since $\mathcal{G}_{\bt_\mu}$ outputs the probability of a spatial point $\bm{x}$ being phase 1, a Sigmoid activation is applied at the output layer to map predictions to the interval $[0,1]$.

\paragraph{The Decoder $\mathcal{G}_{\bt_u}$} The decoder $\mathcal{G}_{\bt_u}$ also uses the MultiONet architecture. Both branch and trunk networks are FFCNs with five hidden layers of $100$ neurons each. Similar to $\mathcal{G}_{\bt_\mu}$, the trunk network applies the $\text{SiLU\_Sin}$ activation, while the branch network applies the $\text{SiLU\_Id}$ activation for all hidden layers.

\paragraph{The Normalizing Flow Model $f_{\bt_\beta}$} The normalizing flow prior consists of three flow steps, each parameterized by a fully connected network with two hidden layers of 64 neurons per layer. The SiLU activation is applied to all hidden layers.

\paragraph{Setups in gradient-based optimization} For all inverse design problems, we adopt a consistent gradient-based optimization setup. Gradients of the objective with respect to $\bm{\beta}$ are computed using PyTorch’s automatic differentiation. The ADAM optimizer is employed with an initial learning rate of $0.01$, which is halved every $100$ steps. A total of $500$ updates are applied to ensure convergence.

\subsection{The PoreFlow framework}
\label{sec:net_cmp}
The PoreFlow framework also employs an encoder $e_{\bm{\eta}}$, a decoder $\mathcal{G}_{\bt_\mu}$, and a conditional normalizing flow (CNF) $f_{\bt_\beta}$, serving roles analogous to those in the proposed Design-GenNO model. Unlike the proposed Design-GenNO, which incorporates governing physics by predicting PDE solutions from latent variables through another decoder $\mathcal{G}_{\bt_T}$, PoreFlow bypasses the physics and directly predicts properties from the latent space through an additional network $\mathcal{G}_{\bt_\kappa}$.

For fair comparison, we adopt the same architectures for $e_{\bm{\eta}}$, $\mathcal{G}_{\bt_\mu}$, and $f_{\bt_\beta}$ as in Design-GenNO, except that the CNF input also includes property values as conditioning variables. The property-prediction network $\mathcal{G}_{\bt_\kappa}$ is modeled as an FFCN with five hidden layers and $100$ neurons per layer, using the $\text{SiLU\_Sin}$ activation \eqref{eq:silu_sin} for all hidden layers.

\end{document}